\documentclass[12pt,a4paper]{article}

\usepackage{cite}
\usepackage{authblk}
\usepackage{amsmath,amssymb}
\usepackage{graphicx}
\usepackage{amsfonts}
\usepackage{epsfig}

\usepackage{tikz}
\usepackage{pgflibrarysnakes}
\usetikzlibrary[snakes]
\usetikzlibrary{arrows,shapes,positioning}
\usetikzlibrary{decorations.markings}
\tikzstyle arrowstyle=[scale=1]
\tikzstyle directed=[postaction={decorate,decoration={markings,
		mark=at position .65 with {\arrow[arrowstyle]{stealth}}}}]
\tikzstyle reverse directed=[postaction={decorate,decoration={markings,
		mark=at position .65 with {\arrowreversed[arrowstyle]{stealth};}}}]

\renewcommand{\baselinestretch}{1.5}

\newcommand{\im}{\mathrm i}

\newcommand{\tr}{\operatorname{Tr}}

\newcommand{\diag}{\operatorname{diag}}
\newcommand{\ket}[1]{\left|#1\right\rangle}      
\newcommand{\bra}[1]{\left\langle #1\right|}     
\newcommand{\eq}{\begin{equation}}
	\newcommand{\en}{\end{equation}}
\newcommand{\bear}{\begin{eqnarray}}
	\newcommand{\ear}{\end{eqnarray}}

\def\weight[#1][#2][#3][#4][#5]{W\!\!
	\left(\!\!\!\left.\begin{array}{cc}
		#4 \!&\! #3\\#1 \!&\! #2\end{array}\!\right|\!\,#5
	\!\right)}

\def\weightT[#1][#2][#3][#4][#5]{\widetilde{W}
	\left(\left.\begin{array}{cc}
		#4&#3\\#1&#2\end{array}\right|\,#5
	\right)}

\title{\mbox{}Correlation functions of the $19$-vertex IRF model and its integrable quantum spin-$1$ chain}
\author{T.S. Tavares\footnote{E-mail: tavares@df.ufscar.br},
	\ H.B. Abreu  \ and \ \ G.A.P. Ribeiro\footnote{E-mail: pavan@df.ufscar.br}}
\affil{Departamento de F\'{i}sica, Universidade Federal de S\~ao Carlos \\ S\~ao Carlos, SP 13565-905, Brazil\vspace{8pt}}

\date{}

\begin{document}
\renewcommand{\baselinestretch}{1.2}
\maketitle
	
\vspace{-.5in}\noindent
\begin{abstract}
We consider the interaction-round-a-face (IRF) version of the nineteen-vertex model (Zamolodchikov-Fateev model) and derive a new quantum spin-$1$ chain, which can be seen as three coupled Zamolodchikov-Fateev spin-$1$ chains with different boundary twists. In the isotropic limit, the associated spin chain reduces to 
three coupled Babujian-Taktahjan spin-$1$ chains with different boundary twists. We compute the ground state short-distance correlation functions of the isotropic model in the thermodynamic limit by exploiting the shared integrable structure between the IRF and vertex models.
\end{abstract}

\thispagestyle{empty}

\newpage

\pagestyle{plain}

\pagenumbering{arabic}
\newpage

\renewcommand{\baselinestretch}{1.5}

\section{Introduction}

The classical bidimensional vertex models and their associated one-dimensional quantum spin chains have a long history. There exist many results concerning spectral properties as well as for correlation functions \cite{BOOK-KBI,BOOK-JM}. In the context of correlation functions, there is a long list of results for quantum spin chains, especially for the $SU(2)$ spin-$1/2$ chain \cite{TAKA,JMMN92,JM96,KMT00,BOKO01,GAS05,BJMST,BST,BGKS06,CORRXXZ,KKMST09,BK2012,MS2018}, its higher-spin realizations \cite{BoWe94,Idzumi94,Kitanine01,DeMa10,GSS10,KNS2013,RK2016,BS2019} and for high-rank spin chains \cite{BOOS2018,RK2019,RIBEIRO2020}.

On the other hand, there are fewer results about correlation functions of the interaction-round-a-face (IRF) models and their associated spin chains \cite{FODA,LUKYANOV,VARCHENKO,NICOLLI2013,TERRAS}. 
The IRF model and its different realizations as the cyclic solid-on-solid (CSOS) \cite{BAXTER1973,BAXTER,BAXTER1984,KUNIBA,PEARCE1988}, the restricted solid-on-solid (RSOS) models and their $A$-$D$-$E$ generalizations \cite{PEARCE1989,PASQUIER,PEARCE1992,PEARCE1993} have some resemblance with the vertex models and its integrable structure, which allows for the exact computation of physical properties. This was exploited recently to devise an approach to deal with correlation functions of face models \cite{FRAHM2021,FRAHM2023}. In \cite{FRAHM2021}, the reduced density matrix was formulated in terms of a face model and the existence of discrete functional equations of qKZ was established. Remarkably, it was shown \cite{FRAHM2021,FRAHM2023} that the density matrix of the RSOS models can be factorized in terms of nearest-neighbour correlators, which allows for the computation of short-distance correlations. In addition, it was also possible to compute the complete density matrix in the ground state of the recently proposed interaction-round-a-face version of the six-vertex model and its associated spin-$1/2$ chain \cite{TAVARES2023,TAVARES2024} for finite and infinite system size.

In this work, we are interested in extending the results \cite{TAVARES2024} to the case of the nineteen-vertex IRF model and its integrable quantum spin-$1$ chain. Here, we construct an IRF model from three copies of the Zamolodchikov-Fateev model\cite{ZF1980}, which is the nineteen-vertex model constructed from the six-vertex model by a fusion procedure \cite{RESHE}. In the Bazhanov \cite{BAZ} and Jimbo \cite{JIMBO} classification, it is the $B^{(1)}_1$ model or the $A^{(1)}_1$ model in the spin-1 representation \cite{RESHE}. The above IRF model results in a new quantum spin-$1$ chain that can be seen as three coupled Zamolodchikov-Fateev spin-$1$ chain \cite{ZF1980} with  a special combination of the boundary twists. Since, in the isotropic limit, the Zamolodchikov-Fateev spin-$1$ chain reduces to the  Babujian-Taktahjian spin-$1$ chain \cite{BATA}, the isotropic limit of our new spin-$1$ is made up of three coupled Babujian-Taktahjian spin-$1$ chains \cite{BATA}. In the isotropic case, we compute the complete density matrix and therefore the correlation function for up to three sites.

This work is organized as follows. In section \ref{integ}, we describe the IRF version of the nineteen-vertex model and its integrable structure.  In section \ref{densitysec}, we introduce the physical density matrix and its functional equation. In section \ref{results}, we present the solution for two- and three-site density matrices of the IRF  model and up to two-site for the spin chain for finite system size and in the thermodynamic limit. In section \ref{finite}, we discuss the computation of the correlation for finite system sizes and compare with the results in the thermodynamic limit. Our conclusions are given in section \ref{CONCLUSION}.

\section{The IRF version of a nineteen-vertex model and its quantum spin-$1$ chain}\label{integ}

Face models are classical statistical mechanical models on a square lattice defined by local Boltzmann weights, which can be depicted as \cite{BAXTER1973,BAXTER,BAXTER1984},
\eq
\begin{aligned}
	\begin{tikzpicture}[scale=1.25]
		\draw (0.0,0.) node {$\weight[a][b][c][d][\lambda]=$};
		\draw (2.0,0) [-,color=black, thick]	+(-0.5,-0.5) -- +(-0.5,0.5)-- +(0.5,0.5) -- +(0.5,-0.5)-- +(-0.5,-0.5);
		\draw (1.45,-0.66) node {$a$};
		\draw (2.55,-0.66) node {$b$};
		\draw (2.55, 0.64) node {$c$};
		\draw (1.45, 0.65) node {$d$};
		\draw (2., 0.) node {$\lambda$};
		\draw (2.,0) [-,color=black,  thick, rounded corners=7pt]	+(-0.3,-0.5) -- +(-0.3,-0.3) -- +(-0.5,-0.3) ;
		\draw (2.75, 0.) node {,};
	\end{tikzpicture}
\end{aligned}
\en
where $a,b,c,d$ are the spins or heights of the corners of the face, separated by bonds and $\lambda$ is the spectral parameter.

In this work, we propose the IRF version of the Zamolodchikov-Fateev model\cite{ZF1980}, which is the nineteen-vertex model constructed from the six-vertex model by a fusion procedure \cite{RESHE}. This vertex model is associated to the Zamolodchikov-Fateev spin chain \cite{ZF1980}.

From the vertex model side, the $R$-matrix, which satisfies the usual Yang-Baxter equation,
\eq
R_{12}(\lambda-\mu)R_{13}(\lambda)R_{23}(\mu)=R_{23}(\mu)R_{13}(\lambda)R_{12}(\lambda-\mu),
\en 
can be written as\cite{ZF1980},
\eq
\!\!\!\!\!R(\lambda)\!=
\!\left(\! \begin{array}{ccc|ccc|ccc}
w_1(\lambda) & 0 & 0 & 0 & 0 & 0 & 0 & 0 & 0 \\
0 & w_2(\lambda) & 0 & w_3(\lambda) & 0 & 0 & 0 & 0 & 0 \\
0 & 0 & w_5(\lambda) & 0 & w_4(\lambda) & 0 & w_6(\lambda) & 0 & 0 \\
\hline
0 & w_3(\lambda) & 0 & w_2(\lambda) & 0 & 0 & 0 & 0 & 0 \\
0 & 0 & w_4(\lambda)  & 0 & w_7(\lambda) & 0 & w_4(\lambda) & 0 & 0 \\
0 & 0 & 0 & 0 & 0 & w_2(\lambda) & 0 & w_3(\lambda) & 0 \\
\hline
0 & 0 & w_6(\lambda) & 0 & w_4(\lambda) & 0 & w_5(\lambda) & 0 & 0 \\
0 & 0 & 0 & 0 & 0 & w_3(\lambda) & 0 & w_2(\lambda) & 0 \\
0 & 0 & 0 & 0 & 0 & 0 & 0 & 6 & w_1(\lambda)
\label{R19v}
\end{array}\right),
\en
where the Boltzmann weights are defined as\cite{ZF1980},
\bear
w_1(\lambda)&=&1, \quad w_2(\lambda)=\frac{\sin(\lambda)}{\sin(\lambda+2\gamma)}, \quad w_3(\lambda)=\frac{\sin(2\gamma)}{\sin(\lambda+2\gamma)}, \quad w_4(\lambda)=\frac{\sin(2 \gamma) \sin(\lambda )}{\sin(\lambda+\gamma)\sin(\lambda+2\gamma)}, \nonumber\\
w_5(\lambda)&=&\frac{\sin(\lambda-\gamma)\sin(\lambda)}{\sin(\lambda+\gamma)\sin(\lambda+2\gamma)}, \quad w_6(\lambda)= \frac{\sin(\gamma) \sin(2 \gamma)}{\sin(\lambda+\gamma)\sin(\lambda+2\gamma)}, \\
w_7(\lambda)&=&-\frac{\left(-2 \cos(\gamma) + \cos(3 \gamma) + \cos(2 \lambda+ \gamma) \right)}{2 \sin(\lambda+ \gamma)\sin(\lambda+2\gamma)}. \nonumber
\ear

On the other hand, the proposed face model can be seen as three copies of the above nineteen-vertex model, whose face weights are given as,
\bear
\weight[a][a\mp1][a][a\pm1][\lambda]&=&w_1(\lambda), 
\nonumber \\
\weight[a][a][a\pm1][a\pm 1][\lambda]&=&\weight[a][a\pm 1][a\pm 1][a][\lambda]=w_2(\lambda),  \nonumber\\
\weight[a][a][a][a\pm1][\lambda]&=&\weight[a][a\pm 1][a][a][\lambda]=w_3(\lambda),\nonumber
\ear
\bear
\weight[a\pm1][a][a][a][\lambda]&=&\weight[a][a][a\pm1][a][\lambda]=w_4(\lambda), \nonumber\\
\weight[a\pm1][a][a\mp1][a][\lambda]&=&w_5(\lambda), \label{wIRF19V} \\
\weight[a][a\pm1][a][a\pm1][\lambda]&=&w_6(\lambda), \nonumber \\
\weight[a][a][a][a][\lambda]&=&w_7(\lambda),\nonumber 
\ear
for $a=0,1,2$. The face version of the nineteen-vertex model is depicted in the Figure \ref{IRF19V}. 

\begin{figure}[thb]
	\begin{minipage}{\linewidth}
		\begin{center}
				
				\begin{tikzpicture}[scale=1.5]

					\draw (0,0) [-,color=black, dotted,  rounded corners=7pt]+(-(0.,0) --+(0.,0.5);
					\draw (0,0) [-,color=black, dotted,  rounded corners=7pt]+(-(0.,-0.5) --+(0.,0);
					\draw (0,0) [-,color=black,  dotted,  rounded corners=7pt]	+(0,0) -- +(0.5,0) ;
					\draw (0,0) [-,color=black,  dotted,  rounded corners=7pt]	+(-0.5,0) -- +(0,0) ;
					\draw (0,0) [-,color=black,  thick, rounded corners=7pt]	+(-0.3,-0.5) -- +(-0.3,-0.3) -- +(-0.5,-0.3) ;
					\draw (0,0) [-,color=black, thick]	+(-0.5,-0.5) -- +(-0.5,0.5)-- +(0.5,0.5) -- +(0.5,-0.5)--+(-0.5,-0.5);
					\draw (-0.55,-0.6) node {\tiny $a$};
					\draw (0.6,-0.6) node {\tiny $a\mp1$};
					\draw (0.6, 0.6) node {\tiny $a$};
					\draw (-0.55, 0.6) node {\tiny $a\pm1$};
					\draw (-0.,-0.6) node {\tiny $2\mp1$};
					\draw (0.75,-0.) node {\tiny $2\mp 1 $};
					\draw (-0.7, 0.) node {\tiny$2\mp 1$};
					\draw (-0., 0.6) node {\tiny$2\mp 1$};
					\draw (-1.5, -0.+0.) node {$w_1(\lambda)=$};
					\draw (1.05, -0.-0.1) node {$,$};
					
					\draw (3,0) [-,color=black, dotted,  rounded corners=7pt]+(-(0.,0) --+(0.,0.5);
					\draw (3,0) [-,color=black, dotted,  rounded corners=7pt]+(-(0.,-0.5) --+(0.,0);
					\draw (3,0) [-,color=black,  dotted,  rounded corners=7pt]	+(0,0) -- +(0.5,0) ;
					\draw (3,0) [-,color=black,  dotted,  rounded corners=7pt]	+(-0.5,0) -- +(0,0) ;
					\draw (3,0) [-,color=black,  thick, rounded corners=7pt]	+(-0.3,-0.5) -- +(-0.3,-0.3) -- +(-0.5,-0.3) ;
					\draw (3,0) [-,color=black, thick]	+(-0.5,-0.5) -- +(-0.5,0.5)-- +(0.5,0.5) -- +(0.5,-0.5)--+(-0.5,-0.5);
					\draw (-0.55+3,-0.6) node {\tiny $a$};
					\draw (0.6+3,-0.6) node {\tiny $a$};
					\draw (0.6+3, 0.6) node {\tiny $a$};
					\draw (-0.55+3, 0.6) node {\tiny $a$};
					\draw (-0.+3,-0.6) node {\tiny $2$};
					\draw (0.6+3,-0.) node {\tiny $2$};
					\draw (-0.6+3, 0.) node {\tiny$2$};
					\draw (-0.+3, 0.6) node {\tiny$2$};
					\draw (1.75, -0.+0.) node {$w_7(\lambda)=$};

\draw (0,-1.5) [-,color=black, dotted,  rounded corners=7pt]+(-(0.,0) --+(0.,0.5);
\draw (0,-1.5) [-,color=black, dotted,  rounded corners=7pt]+(-(0.,-0.5) --+(0.,0);
\draw (0,-1.5) [-,color=black,  dotted,  rounded corners=7pt]	+(0,0) -- +(0.5,0) ;
\draw (0,-1.5) [-,color=black,  dotted,  rounded corners=7pt]	+(-0.5,0) -- +(0,0) ;
\draw (0,-1.5) [-,color=black,  thick, rounded corners=7pt]	+(-0.3,-0.5) -- +(-0.3,-0.3) -- +(-0.5,-0.3) ;
\draw (0,-1.5) [-,color=black, thick]	+(-0.5,-0.5) -- +(-0.5,0.5)-- +(0.5,0.5) -- +(0.5,-0.5)--+(-0.5,-0.5);
\draw (-0.55,-0.6-1.5) node {\tiny $a$};
\draw (0.6,-0.6-1.5) node {\tiny $a$};
\draw (0.6, 0.6-1.5) node {\tiny $a\pm 1$};
\draw (-0.55, 0.6-1.5) node {\tiny $a\pm 1$};
\draw (-0.,-0.6-1.5) node {\tiny $2$};
\draw (0.7,-0.-1.5) node {\tiny $2\mp 1$};
\draw (-0.7, 0.-1.5) node {\tiny$2 \mp 1$};
\draw (-0., 0.6-1.5) node {\tiny$2$};
\draw (-1.5, -0.+0.-1.5) node {$w_2(\lambda)=$};

\draw (2,-1.5) [-,color=black, dotted,  rounded corners=7pt]+(-(0.,0) --+(0.,0.5);
\draw (2,0-1.5) [-,color=black, dotted,  rounded corners=7pt]+(-(0.,-0.5) --+(0.,0);
\draw (2,0-1.5) [-,color=black,  dotted,  rounded corners=7pt]	+(0,0) -- +(0.5,0) ;
\draw (2,0-1.5) [-,color=black,  dotted,  rounded corners=7pt]	+(-0.5,0) -- +(0,0) ;
\draw (2,0-1.5) [-,color=black,  thick, rounded corners=7pt]	+(-0.3,-0.5) -- +(-0.3,-0.3) -- +(-0.5,-0.3) ;
\draw (2,0-1.5) [-,color=black, thick]	+(-0.5,-0.5) -- +(-0.5,0.5)-- +(0.5,0.5) -- +(0.5,-0.5)--+(-0.5,-0.5);
\draw (-0.55+2,-0.6-1.5) node {\tiny $a$};
\draw (0.6+2,-0.6-1.5) node {\tiny $a\pm 1$};
\draw (0.6+2, 0.6-1.5) node {\tiny $a\pm 1$};
\draw (-0.55+2, 0.6-1.5) node {\tiny $a$};
\draw (-0.+2,-0.6-1.5) node {\tiny $2\pm 1$};
\draw (0.6+2,-0.-1.5) node {\tiny $2$};
\draw (-0.6+2, 0.-1.5) node {\tiny$2$};
\draw (-0.+2, 0.6-1.5) node {\tiny$2\pm 1$};
\draw (1.1,-0.-1.5) node {\tiny $=$};

\draw (0,-3) [-,color=black, dotted,  rounded corners=7pt]+(-(0.,0) --+(0.,0.5);
\draw (0,-3) [-,color=black, dotted,  rounded corners=7pt]+(-(0.,-0.5) --+(0.,0);
\draw (0,-3) [-,color=black,  dotted,  rounded corners=7pt]	+(0,0) -- +(0.5,0) ;
\draw (0,-3) [-,color=black,  dotted,  rounded corners=7pt]	+(-0.5,0) -- +(0,0) ;
\draw (0,-3) [-,color=black,  thick, rounded corners=7pt]	+(-0.3,-0.5) -- +(-0.3,-0.3) -- +(-0.5,-0.3) ;
\draw (0,-3) [-,color=black, thick]	+(-0.5,-0.5) -- +(-0.5,0.5)-- +(0.5,0.5) -- +(0.5,-0.5)--+(-0.5,-0.5);
\draw (-0.55,-0.6-3) node {\tiny $a$};
\draw (0.6,-0.6-3) node {\tiny $a$};
\draw (0.6, 0.6-3) node {\tiny $a$};
\draw (-0.55, 0.6-3) node {\tiny $a\pm 1$};
\draw (-0.,-0.6-3) node {\tiny $2$};
\draw (0.7,-0.-3) node {\tiny $2$};
\draw (-0.7, 0.-3) node {\tiny$2 \mp 1$};
\draw (-0., 0.6-3) node {\tiny$2\mp1$};
\draw (-1.5, -0.+0.-3.0) node {$w_3(\lambda)=$};

\draw (2,-3.0) [-,color=black, dotted,  rounded corners=7pt]+(-(0.,0) --+(0.,0.5);
\draw (2,0-3.0) [-,color=black, dotted,  rounded corners=7pt]+(-(0.,-0.5) --+(0.,0);
\draw (2,0-3.) [-,color=black,  dotted,  rounded corners=7pt]	+(0,0) -- +(0.5,0) ;
\draw (2,0-3.) [-,color=black,  dotted,  rounded corners=7pt]	+(-0.5,0) -- +(0,0) ;
\draw (2,0-3.) [-,color=black,  thick, rounded corners=7pt]	+(-0.3,-0.5) -- +(-0.3,-0.3) -- +(-0.5,-0.3) ;
\draw (2,0-3.) [-,color=black, thick]	+(-0.5,-0.5) -- +(-0.5,0.5)-- +(0.5,0.5) -- +(0.5,-0.5)--+(-0.5,-0.5);
\draw (-0.55+2,-0.6-3.) node {\tiny $a$};
\draw (0.6+2,-0.6-3.) node {\tiny $a\pm 1$};
\draw (0.6+2, 0.6-3.) node {\tiny $a$};
\draw (-0.55+2, 0.6-3.) node {\tiny $a$};
\draw (-0.+2,-0.6-3.) node {\tiny $2\pm 1$};
\draw (0.7+2,-0.-3.) node {\tiny $2\pm1$};
\draw (-0.6+2, 0.-3.) node {\tiny$2$};
\draw (-0.+2, 0.6-3.) node {\tiny$2$};
\draw (1.05,-0.-3.) node {\tiny $=$};

\draw (0,-4.5) [-,color=black, dotted,  rounded corners=7pt]+(-(0.,0) --+(0.,0.5);
\draw (0,-4.5) [-,color=black, dotted,  rounded corners=7pt]+(-(0.,-0.5) --+(0.,0);
\draw (0,-4.5) [-,color=black,  dotted,  rounded corners=7pt]	+(0,0) -- +(0.5,0) ;
\draw (0,-4.5) [-,color=black,  dotted,  rounded corners=7pt]	+(-0.5,0) -- +(0,0) ;
\draw (0,-4.5) [-,color=black,  thick, rounded corners=7pt]	+(-0.3,-0.5) -- +(-0.3,-0.3) -- +(-0.5,-0.3) ;
\draw (0,-4.5) [-,color=black, thick]	+(-0.5,-0.5) -- +(-0.5,0.5)-- +(0.5,0.5) -- +(0.5,-0.5)--+(-0.5,-0.5);
\draw (-0.55,-0.6-4.5) node {\tiny $a\pm 1$};
\draw (0.6,-0.6-4.5) node {\tiny $a$};
\draw (0.6, 0.6-4.5) node {\tiny $a$};
\draw (-0.55, 0.6-4.5) node {\tiny $a$};
\draw (-0.,-0.6-4.5) node {\tiny $2\mp1$};
\draw (0.7,-0.-4.5) node {\tiny $2$};
\draw (-0.7, 0.-4.5) node {\tiny$2 \pm 1$};
\draw (-0., 0.6-4.5) node {\tiny$2$};
\draw (-1.5, -0.+0.-4.5) node {$w_4(\lambda)=$};

\draw (2,-4.5) [-,color=black, dotted,  rounded corners=7pt]+(-(0.,0) --+(0.,0.5);
\draw (2,0-4.5) [-,color=black, dotted,  rounded corners=7pt]+(-(0.,-0.5) --+(0.,0);
\draw (2,0-4.5) [-,color=black,  dotted,  rounded corners=7pt]	+(0,0) -- +(0.5,0) ;
\draw (2,0-4.5) [-,color=black,  dotted,  rounded corners=7pt]	+(-0.5,0) -- +(0,0) ;
\draw (2,0-4.5) [-,color=black,  thick, rounded corners=7pt]	+(-0.3,-0.5) -- +(-0.3,-0.3) -- +(-0.5,-0.3) ;
\draw (2,0-4.5) [-,color=black, thick]	+(-0.5,-0.5) -- +(-0.5,0.5)-- +(0.5,0.5) -- +(0.5,-0.5)--+(-0.5,-0.5);
\draw (-0.55+2,-0.6-4.5) node {\tiny $a$};
\draw (0.6+2,-0.6-4.5) node {\tiny $a$};
\draw (0.6+2, 0.6-4.5) node {\tiny $a\pm 1$};
\draw (-0.55+2, 0.6-4.5) node {\tiny $a$};
\draw (-0.+2,-0.6-4.5) node {\tiny $2$};
\draw (0.71+2,-0.-4.5) node {\tiny $2\mp1$};
\draw (-0.6+2, 0.-4.5) node {\tiny$2$};
\draw (-0.+2, 0.6-4.5) node {\tiny$2\pm1$};
\draw (1.05,-0.-4.5) node {\tiny $=$};

\draw (0,-6) [-,color=black, dotted,  rounded corners=7pt]+(-(0.,0) --+(0.,0.5);
\draw (0,-6) [-,color=black, dotted,  rounded corners=7pt]+(-(0.,-0.5) --+(0.,0);
\draw (0,-6) [-,color=black,  dotted,  rounded corners=7pt]	+(0,0) -- +(0.5,0) ;
\draw (0,-6) [-,color=black,  dotted,  rounded corners=7pt]	+(-0.5,0) -- +(0,0) ;
\draw (0,-6) [-,color=black,  thick, rounded corners=7pt]	+(-0.3,-0.5) -- +(-0.3,-0.3) -- +(-0.5,-0.3) ;
\draw (0,-6) [-,color=black, thick]	+(-0.5,-0.5) -- +(-0.5,0.5)-- +(0.5,0.5) -- +(0.5,-0.5)--+(-0.5,-0.5);
\draw (-0.55,-0.6-6) node {\tiny $a\pm1$};
\draw (0.6,-0.6-6) node {\tiny $a$};
\draw (0.6, 0.6-6) node {\tiny $a\mp1$};
\draw (-0.55, 0.6-6) node {\tiny $a$};
\draw (-0.,-0.6-6) node {\tiny $2\mp1$};
\draw (0.75,-0.-6) node {\tiny $2\pm 1 $};
\draw (-0.7, 0.-6) node {\tiny$2\pm 1$};
\draw (-0., 0.6-6) node {\tiny$2\mp 1$};
\draw (-1.5, -0.+0.-6) node {$w_5(\lambda)=$};
\draw (1.05, -0.-0.1-6) node {$,$};

\draw (3,0-6) [-,color=black, dotted,  rounded corners=7pt]+(-(0.,0) --+(0.,0.5);
\draw (3,0-6) [-,color=black, dotted,  rounded corners=7pt]+(-(0.,-0.5) --+(0.,0);
\draw (3,0-6) [-,color=black,  dotted,  rounded corners=7pt]	+(0,0) -- +(0.5,0) ;
\draw (3,0-6) [-,color=black,  dotted,  rounded corners=7pt]	+(-0.5,0) -- +(0,0) ;
\draw (3,0-6) [-,color=black,  thick, rounded corners=7pt]	+(-0.3,-0.5) -- +(-0.3,-0.3) -- +(-0.5,-0.3) ;
\draw (3,0-6) [-,color=black, thick]	+(-0.5,-0.5) -- +(-0.5,0.5)-- +(0.5,0.5) -- +(0.5,-0.5)--+(-0.5,-0.5);
\draw (-0.55+3,-0.6-6) node {\tiny $a$};
\draw (0.6+3,-0.6-6) node {\tiny $a\pm1$};
\draw (0.6+3, 0.6-6) node {\tiny $a$};
\draw (-0.55+3, 0.6-6) node {\tiny $a\pm1$};
\draw (-0.+3,-0.6-6) node {\tiny $2\pm1$};
\draw (0.71+3,-0.-6) node {\tiny $2\pm1$};
\draw (-0.7+3, 0.-6) node {\tiny$2\mp 1$};
\draw (-0.+3, 0.6-6) node {\tiny$2\mp1$};
\draw (1.6, -0.+0.-6) node {$w_6(\lambda)=$};

				\end{tikzpicture}

		\end{center}
	\end{minipage}
	\caption{Boltzmann weights of the face model, which are obtained from the allowed configuration for the nineteen-vertex model indicated by the dotted lines at the center.}
	\label{IRF19V}
\end{figure}

This construction guarantees that the above face weights satisfy the face version of the Yang-Baxter equation,
\bear
\sum_{i} \weight[a][b][i][f][\lambda-\mu]\weight[i][d][e][f][\mu]\weight[b][c][d][i][\lambda]\nonumber\\
=\sum_{i} \weight[a][i][e][f][\lambda]\weight[b][c][i][a][\mu]\weight[i][c][d][e][\lambda-\mu].
\ear

The Yang-Baxter equation implies into the commutation of the IRF transfer matrix ($[T_{\text{IRF}}(\lambda),T_{\text{IRF}}(\mu)]=0$), which can be defined as,
\eq
T_{\text{IRF}}(\lambda)=\tr[{\cal T}(\lambda)]=\sum_{\alpha_1,\alpha_2} {\cal T}^{\alpha_1 \alpha_1}_{\alpha_2\alpha_2}(\lambda),
\en 
where the monodromy matrix ${\cal T}(\lambda)$ is an ordered product of the $L$ Boltzmann weights,
\bear
{{\cal T}^{\alpha_1\beta_1}_{\alpha_2 \beta_2}}(\lambda)^{a_1 a_2 \cdots a_L}_{b_1 b_2 \cdots b_L}=\prod_{i=1}^L \weight[b_{i}][b_{i+1}][a_{i+1}][a_{i}][\lambda-u_i]\delta_{\alpha_1 a_1}\delta_{\alpha_2 b_1}\delta_{\beta_1 a_{L+1}}\delta_{\beta_2 b_{L+1}}.
\label{monodromy}
\ear
which can be depicted as in Figure \ref{monodromy-figure}.
\begin{figure}[h]
	\begin{minipage}{\linewidth}
		\begin{center}
			\begin{tikzpicture}[scale=1.5]
				\draw (0.45,-0.65) node {$\alpha_2=b_1$};
\draw (1.55,-0.65) node {$b_2$};
\draw (2.55,-0.65) node {$b_3$};
\draw (3.55,-0.65) node {$b_4$};
\draw (4.55,-0.65) node {$b_5$};
\draw (5.65,-0.65) node {$b_{L}$};
\draw (6.75,-0.66) node {$b_{L+1}=\beta_2$};
\draw (0.45, 0.75) node {$\alpha_1=a_1$};
\draw (1.55,0.75) node {$a_2$};
\draw (2.55,0.75) node {$a_3$};
\draw (3.55,0.75) node {$a_4$};
\draw (4.55,0.75) node {$a_5$};
\draw (5.65,0.75) node {$a_{L}$};
\draw (6.75, 0.75) node {$a_{L+1}=\beta_1$};
				\draw (1.0,0) [-,color=black, thick]	+(-0.5,-0.5) -- +(-0.5,0.5)-- +(0.5,0.5) -- +(0.5,-0.5)-- +(-0.5,-0.5);
				\draw (1.01,0.01) [-,color=black,  rounded corners=7pt,thick]	+(-0.35,-0.5) -- +(-0.35,-0.35)-- +(-0.5,-0.35);
				\draw (2.0,0) [-,color=black, thick]	+(-0.5,-0.5) -- +(-0.5,0.5)-- +(0.5,0.5) -- +(0.5,-0.5)-- +(-0.5,-0.5);
				\draw (2.01,0.01) [-,color=black,  rounded corners=7pt,thick]	+(-0.35,-0.5) -- +(-0.35,-0.35)-- +(-0.5,-0.35);
				\draw (3.0,0) [-,color=black, thick]	+(-0.5,-0.5) -- +(-0.5,0.5)-- +(0.5,0.5) -- +(0.5,-0.5)-- +(-0.5,-0.5);
				\draw (3.01,0.01) [-,color=black,  rounded corners=7pt,thick]	+(-0.35,-0.5) -- +(-0.35,-0.35)-- +(-0.5,-0.35);
				\draw (4.0,0) [-,color=black, thick]	+(-0.5,-0.5) -- +(-0.5,0.5)-- +(0.5,0.5) -- +(0.5,-0.5)-- +(-0.5,-0.5);
				\draw (4.01,0.01) [-,color=black,  rounded corners=7pt,thick]	+(-0.35,-0.5) -- +(-0.35,-0.35)-- +(-0.5,-0.35);
				\draw (5.0,0) [-,color=black, thick]	+(-0.5,-0.5) -- +(-0.5,0.5)-- +(0.5,0.5) -- +(0.5,-0.5)-- +(-0.5,-0.5);
				\draw (6.0,0) [-,color=black, thick]	+(-0.5,-0.5) -- +(-0.5,0.5)-- +(0.5,0.5) -- +(0.5,-0.5)-- +(-0.5,-0.5);
				\draw (6.01,0.01) [-,color=black,  rounded corners=7pt,thick]	+(-0.35,-0.5) -- +(-0.35,-0.35)-- +(-0.5,-0.35);
				\draw (-0.650,0.) node {${\cal T}^{\alpha_1\beta_1}_{\alpha_2 \beta_2}(\lambda)^{a_1 a_2 \cdots a_L}_{b_1 b_2 \cdots b_L}=$};

				\draw (1., 0.) node {$\lambda-u_1$};
				\draw (2., 0.) node {$\lambda-u_2$};
				\draw (3., 0.) node {$\lambda-u_3$};
				\draw (4., 0.) node {$\lambda-u_4$};
				\draw (6., 0.) node {$\lambda-u_L$};
				\draw (5, 0.) node {$\cdots$};
			\end{tikzpicture}
		\end{center}
	\end{minipage}
	\caption{The depiction of the monodromy matrix elements of the IRF model.}
	\label{monodromy-figure}
\end{figure}

The one-dimensional quantum spin chain is obtained by the logarithmic derivative of the IRF transfer matrix ${\cal H}_{\text{IRF}}=\sin(2\gamma) \partial_{\lambda}\log T_{\text{IRF}}( \lambda)|_{\lambda=u_i=0}$, where in this case one has three spin interaction Hamiltonian with periodic boundary condition,
\begin{equation}
{\cal H}_{\text{IRF}}=\sum_{i=1}^L h_{i,i+1,i+2}, \label{Hirf}
\end{equation}
\begin{multline}
h_{1,2,3} = (2-4 \Delta^2)+(2 \Delta^2-1) \left({(S_1^z)}^2+{(S_3^z)}^2\right)-{(S_2^z)}^2+\left(\frac{1}{2}-2 \Delta^2\right) \left(S_1^z S_3^z+{(S_1^z S_3^z)}^2 \right)\\+ \left(\frac{1}{2}-\Delta^2\right) \left(S_1^z S_2^z+S_2^z S_3^z\right)
+\left(\frac{1}{2}+\Delta^2\right) \left({\left(S_1^z S_2^z\right)}^2+{\left(S_2^z S_3^z\right)}^2\right)+\\ \left(\Delta^2-\frac{1}{4}\right) \left(S_1^z S_2^z {(S_3^z)}^2+S_1^z{(S_2^z)}^2 S_3^z +{(S_1^z)}^2 S_2^z S_3^z \right)+
3\left(\frac{1}{4}- \Delta^2\right){(S_1^z S_2^z S_3^z)}^2 + 2 \sqrt{2} \Delta S_2^x\\
  +\left({(S_1^z)}^2+{(S_3^z)}^2-1\right) \left({(S_2^x)}^2-{(S_2^y)}^2\right)+\left(\Delta-\frac{1}{2}\right)\left(S_1^z S_3^z+{(S_1^z S_3^z)}^2 \right) \left({(S_2^x)}^2-{(S_2^y)}^2\right)\\
+\frac{1}{\sqrt{2}} \left(S_1^z+S_3^z\right)  {\left[S_2^z,S_2^x\right]}_+  + \left(\frac{2 \Delta-1}{2 \sqrt{2}}\right)  \left({\left(S_1^z\right)}^2 S_3^z+S_1^z{\left(S_3^z\right)}^2 \right) {\left[S_2^z,S_2^x\right]}_+ + \\ 
\left(\frac{1-4 \Delta}{\sqrt{2}}\right) \left({(S_1^z)}^2+{(S_3^z)}^2\right) S_2^x+\left(\frac{2 \Delta-1}{2\sqrt{2}}\right) S_1^z S_3^z S_2^x+\left(\frac{5(2 \Delta-1)}{2\sqrt{2}}\right) {\left(S_1^z S_3^z\right)}^2 S_2^x, 
\label{Hirf2}
\end{multline} 
where $S^{\alpha}$ for $\alpha=x,y,z,$ are the standard spin-$1$ matrices, $[A,B]_+=A.B + B.A$ is the anticommutator and $\Delta=\cos(\gamma)$. The Hamiltonian (\ref{Hirf}-\ref{Hirf2}) has a continuous $U(1)$ symmetry and a discrete $\mathbb{Z}_3$ symmetry, since it commutes with the operators,
\eq
\Sigma^z= \sum_{i=1}^L  \frac{1}{2} \left(S_{i+1}^z-S_i^z\right) \left(1+S_{i}^z S_{i+1}^z\right), \qquad \Pi^x=\prod_{i=1}^L \left(\begin{array}{ccc}
	0 & 1 & 0\\ 0 & 0& 1\\ 1 & 0 &0
\end{array}\right)_i. \label{symmetries}
\en

The Hamiltonian (\ref{Hirf}-\ref{Hirf2}) can be seen as three coupled 
Zamolodchikov-Fateev (ZF) spin chains \cite{ZF1980} with different boundary conditions restricted to the sector $0 ~\mbox{mod}~3$, i.e. the allowed states correspond to magnetic sectors of spin-$z$ that are multiple of 3,
\eq
{\cal H}_{\text{IRF}} = U^{-1} \left( {\cal H}_{\text{ZF}}^{\text{mod(3)}}(0) \oplus {\cal H}_{\text{ZF}}^{\text{mod(3)}}(2\pi/3) \oplus {\cal H}_{\text{ZF}}^{\text{mod(3)}}(4\pi/3) \right)U, \label{MAP}
\en
where $U$ is the matrix that diagonalizes $\Pi^x$ and $\phi$ is the twist angle. The Zamolodchikov-Fateev spin-$1$ chains ${\cal H}_{ZF}(\phi)$ associated to the vertex model can be written as,  
\bear
{\cal H}_{ZF}(\phi)&=&\sum_{i=1}^{L-1} \hat{h}_{i,i+1} + {\cal G}_L^{-1} \cdot \hat{h}_{L,1} \cdot {\cal G}_L, \\
\hat{h}_{i,j} &=& \vec{S}_{i} \cdot \vec{S}_{j}-(\vec{S}_{i} \cdot \vec{S}_{j})^2  + 2(\Delta^2-1) \left(S^z_{i} S^z_{j} -(S^z_{i} S^z_{j})^2  + (S^z_{i})^2 + (S^z_{j})^2\right) \nonumber  \\
&-&2 (\Delta-1) [S^x_{i} S^x_{j} + S^y_{i} S^y_{j},  S^z_{i} S^z_{j}]_{+}  + 2(\Delta^2-1) I, 
\ear
where $I$ is the identity matrix and the twist angle $\phi$ enters through the diagonal matrix ${\cal G}\!=\!\diag(\exp(\im  \phi),1,\exp(-\im  \phi))$. The value $\phi=0$ refers to periodic boundary condition.

In particular, the isotropic limit $\gamma\to 0$ (or $\Delta\to1$)  , results in the spin-$1$ Babujian-Taktahjian chain, ${\cal H}_{BT}=\lim_{\Delta\to 1}{\cal H}_{ZF}$. On the other hand, at the vertex model level, the isotropic limit of the $R$-matrix (\ref{R19v}) is obtained after scaling the spectral parameter $\lambda \to \gamma\lambda $ and taking the limit $\gamma\to 0$.
From now on, we consider only the isotropic case ($\Delta=1$) to proceed with the computation of the correlation functions.

\section{Density matrix}\label{densitysec}

In \cite{GoKlSe04} it was developed a framework to address the problem of thermal correlation functions of quantum integrable spin chains associated to vertex models. This approach was recently extended \cite{FRAHM2021} to deal with correlation functions of integrable models associated to face models, which allowed for the computation of correlation functions for the RSOS model \cite{FRAHM2023} and the IRF version of the six-vertex model \cite{TAVARES2024}. In this approach, the inhomogeneous reduced density matrix plays an important role. The reduced density matrix, in the context of face models at zero temperature and finite system size $L$, can be written as (see Figure \ref{RDM}),
\eq
D_n(\lambda_1,\lambda_2,\dots,\lambda_n)^{\alpha_1\alpha_2\dots\alpha_{n+1}}_{\beta_1\beta_2\dots\beta_{n+1}}=\frac{\bra{\Phi_0} {\cal T}^{\alpha_1\beta_1}_{\alpha_2\beta_2}(\lambda_1) {\cal T}^{\alpha_2\beta_2}_{\alpha_3\beta_3}(\lambda_2)\dots {\cal T}^{\alpha_n\beta_n}_{\alpha_{n+1}\beta_{n+1}}(\lambda_n) \ket{\Phi_0}}{\Lambda_0(\lambda_1)\Lambda_0(\lambda_2)\cdots \Lambda_0(\lambda_n)},
\label{density}
\en
where ${{\cal T}^{\alpha_k\beta_k}_{\alpha_{k+1} \beta_{k+1}}}(\lambda)$ are the monodromy matrix elements (\ref{monodromy}), $\Lambda_0(x)$ is the  leading eigenvalue associated to the normalized leading eigenstate $\ket{\Phi_0}$ of the IRF transfer matrix $T_{\text{IRF}}(\lambda)$. The physical density matrix of the face model for $n$ consecutive edges in the ground state is obtained in the homogeneous limit $\lambda_k=u_k \to 0$ such that $D_n(0,0,\cdots,0)$  \cite{FRAHM2021,TAVARES2024}.

\begin{figure}[h]
	\begin{minipage}{\linewidth}
		\begin{center}
			\begin{tikzpicture}[scale=1.45]
				\draw (1.0,2) [-,color=black, thick]	+(-0.5,0.5) -- +(1.5,1.25)-- +(3.5,0.5);
				\draw (1.0,0) [-,color=black, thick]	+(-0.5,-0.5) -- +(1.5,-1.25)-- +(3.5,-0.5);
				
				\draw (1.0,2) [-,color=black, thick]	+(-0.5,-0.5) -- +(-0.5,0.5)-- +(0.5,0.5) -- +(0.5,-0.5)-- +(-0.5,-0.5);
				\draw (2.0,2) [-,color=black, thick]	+(-0.5,-0.5) -- +(-0.5,0.5)-- +(0.5,0.5) -- +(0.5,-0.5)-- +(-0.5,-0.5);
				\draw (3.0,2) [-,color=black, thick]	+(-0.5,-0.5) -- +(-0.5,0.5)-- +(0.5,0.5) -- +(0.5,-0.5)-- +(-0.5,-0.5);
				\draw (4.0,2) [-,color=black, thick]	+(-0.5,-0.5) -- +(-0.5,0.5)-- +(0.5,0.5) -- +(0.5,-0.5)-- +(-0.5,-0.5);
				
				\draw (1.0,1) [-,color=black, thick]	+(-0.5,-0.5) -- +(-0.5,0.5)-- +(0.5,0.5) -- +(0.5,-0.5)-- +(-0.5,-0.5);
				\draw (2.0,1) [-,color=black, thick]	+(-0.5,-0.5) -- +(-0.5,0.5)-- +(0.5,0.5) -- +(0.5,-0.5)-- +(-0.5,-0.5);
				\draw (3.0,1) [-,color=black, thick]	+(-0.5,-0.5) -- +(-0.5,0.5)-- +(0.5,0.5) -- +(0.5,-0.5)-- +(-0.5,-0.5);
				\draw (4.0,1) [-,color=black, thick]	+(-0.5,-0.5) -- +(-0.5,0.5)-- +(0.5,0.5) -- +(0.5,-0.5)-- +(-0.5,-0.5);
				
				\draw (1.0,0) [-,color=black, thick]	+(-0.5,-0.5) -- +(-0.5,0.5)-- +(0.5,0.5) -- +(0.5,-0.5)-- +(-0.5,-0.5);
				\draw (2.0,0) [-,color=black, thick]	+(-0.5,-0.5) -- +(-0.5,0.5)-- +(0.5,0.5) -- +(0.5,-0.5)-- +(-0.5,-0.5);
				\draw (3.0,0) [-,color=black, thick]	+(-0.5,-0.5) -- +(-0.5,0.5)-- +(0.5,0.5) -- +(0.5,-0.5)-- +(-0.5,-0.5);
				\draw (4.0,0) [-,color=black, thick]	+(-0.5,-0.5) -- +(-0.5,0.5)-- +(0.5,0.5) -- +(0.5,-0.5)-- +(-0.5,-0.5);
				
				\draw (2.55, 2.85) node {$\Phi_0$};
				\draw (2.55, -0.85) node {$\Phi_0$};				
				
				\draw (0.3, 1.5) node {$\alpha_2$};
				\draw (4.7, 1.5) node {$\beta_2$};
				\draw (0.3, 2.5) node {$\alpha_1$};
				\draw (4.7, 2.5) node {$\beta_1$};
				
				\draw (0.2,-0.5) node {$\alpha_{n+1}$};
				\draw (4.8,-0.5) node {$\beta_{n+1}$};
				\draw (0.3, 0.5) node {$\alpha_n$};
				\draw (4.7, 0.5) node {$\beta_n$};
				
				\draw (-1.50,1.) node {$D_n(\lambda_1,\dots,\lambda_n)^{\alpha_1\alpha_2\dots\alpha_{n+1}}_{\beta_1\beta_2\dots\beta_{n+1}}=$};
				
				\draw (1., 2.) node {$\lambda_1-u_1$};
				\draw (2., 2.) node {$\lambda_1-u_2$};
				\draw (4., 2.) node {$\lambda_1-u_L$};
				\draw (3, 2.) node {$\cdots$};
				\draw (1., 1.) node {$\vdots$};
				\draw (4., 1.) node {$\vdots$};
				\draw (3, 0.) node {$\cdots$};
				\draw (1., 0.) node {$\lambda_n-u_1$};
				\draw (2., 0.) node {$\lambda_n-u_2$};
				\draw (4., 0.) node {$\lambda_n-u_L$};
				\draw (3, 0.) node {$\cdots$};
				
			\end{tikzpicture}
		\end{center}
	\end{minipage}
	\caption{Graphical depiction of the reduced density matrix.}
	\label{RDM}
\end{figure}

The density matrix of the associated quantum spin chain (${\mathbb D}_{n-1}$) is obtained from the partial trace of the IRF density matrix, which results in,
\eq
{\mathbb D}_{n-1}(\lambda_1,\lambda_2,\cdots,\lambda_n)^{\alpha_2\dots\alpha_{n}}_{\beta_2\dots\beta_{n}} =\sum_{\alpha_1,\alpha_{n+1}} D_n(\lambda_1,\lambda_2,\cdots,\lambda_n)^{\alpha_1\alpha_2\dots\alpha_n\alpha_{n+1}}_{\alpha_1\beta_2\dots\beta_n\alpha_{n+1}}.
\en

The computation of the inhomogeneous reduced density matrices can be done through the solution of a discrete functional equation of qKZ type \cite{BK2012,FRAHM2021,TAVARES2024}. The existence of such equation is guaranteed by the integrable structure and the crossing symmetry of the Boltzmann weights. This structure also holds true for the case of higher-spin chain \cite{KNS2013}.

The face version of the discrete functional equations is given by  \cite{FRAHM2021},
\eq
D_n(\lambda_1,\lambda_2,\cdots,\lambda_n-1)= A_n(\lambda_1,\lambda_2,\cdots,\lambda_n)[D_n(\lambda_1,\lambda_2,\cdots,\lambda_n)], \label{qKZ}
\en
where the linear operator $A_n$ can be written as,
\bear
&A_n(\lambda_1,\lambda_2,\cdots,\lambda_n)[B]^{\alpha_1\alpha_2\dots\alpha_{n+1}}_{\beta_1\beta_2\dots\beta_{n+1}}=\frac{\delta_{\alpha_1 \beta_1}\delta_{\alpha_{n+1} \beta_{n+1}}}{2\prod_{k=1}^n (1-(\lambda_k-\lambda_n)^2)(4-(\lambda_k-\lambda_n)^2)} \times \nonumber \\
&\displaystyle\sum_{\gamma_i,\delta_i=\pm} \delta_{\alpha_n\gamma_{n+1}} \prod_{k=1}^{n-1}\weight[\alpha_k][\alpha_{k+1}][\gamma_{k+1}][\gamma_{k}][\lambda_n-\lambda_k] B^{\gamma_1\dots\gamma_{n+1}}_{\delta_1\dots\delta_{n+1}} \\
&\prod_{k=1}^{n-1}\weight[\delta_{k+1}][\beta_{k+1}][\beta_{k}][\delta_{k}][\lambda_k-\lambda_n] \weight[\delta_{n+1}][\beta_{n+1}][\beta_{n}][\delta_{n}][-1 ]. \nonumber
\ear
The above discrete functional equations (\ref{qKZ}) were verified by direct inspection for small lattices.

In the next section, we are going to propose a suitable ansatz for the density matrix which allows for the  solution of the functional equation for short-distances $n=2$ and $3$ for finite lattice sizes and in the thermodynamic limit.

\section{Computation of the reduced density matrix}\label{results}

Due to the fact that the IRF model (\ref{wIRF19V}) is made of three copies of the nineteen model, and likewise the associated spin chain (\ref{Hirf}) in the isotropic limit is made of three coupled Babujian-Taktahjian spin-$1$ chains,
the IRF reduced density matrix in the ground state (\ref{density}) can be written, in a proper basis, as a direct sum of three copies of the density matrix of the Babujian-Taktahjian spin-$1$ chains,
\eq
D_n^{\text{IRF}}(\lambda_1,\dots,\lambda_n)= \frac{1}{3} D_n^{BT}(\lambda_1,\dots,\lambda_n) \oplus  \frac{1}{3} D_n^{BT}(\lambda_1,\dots,\lambda_n)\oplus  \frac{1}{3} D_n^{BT}(\lambda_1,\dots,\lambda_n).
\en

The computation of the two and three-site density matrix for the  Babujian-Taktahjian spin-$1$ chains was performed in \cite{KNS2013}, therefore, we can exploit the shared integrable structure between the face and vertex model, to evaluate the IRF density matrix and likewise for its associated spin chain.

\subsection{Computation of the two-site density matrix}\label{2siteDM}

The two-site density matrix of the isotropic spin-$1$ chain can be written in terms of the identity, permutation and Termperley-Lieb operators \cite{KNS2013}, such that
\eq
D_2^{BT}(\lambda_1,\lambda_2)=\rho_1(\lambda_1,\lambda_2)I_9 +  \rho_2(\lambda_1,\lambda_2) P_{12} + \rho_3(\lambda_1,\lambda_2) E_{12},
\label{D2XXX}
\en
where due to the density matrix normalization   $\rho_1(\lambda_1,\lambda_2)=\tfrac{1}{9}-\tfrac{1}{3}\rho_2(\lambda_1,\lambda_2)-\tfrac{1}{3}\rho_3(\lambda_1,\lambda_2)$. This is due to the fact the in the isotropic limit the $R$-matrix (\ref{R19v}) can be written explicitly as $R_{12}(\lambda)=\lambda I + P_{12}- \frac{\lambda}{\lambda+\tfrac{1}{2}} E_{12}$. Besides one has that  $I_9$ is the $9\times9$ identity and $P_{12}$ and $E_{12}$ are the permutation and monoid operators,
\begin{align}
P_{12}\!=\!\left(\!\begin{array}{ccc|ccc|ccc}
1 & 0 & 0 & 0 & 0 & 0 & 0 & 0 & 0 \\
0 & 0 & 0 & 1 & 0 & 0 & 0 & 0 & 0 \\
0 & 0 & 0 & 0 & 0 & 0 & 1 & 0 & 0 \\
\hline
0 & 1 & 0 & 0 & 0 & 0 & 0 & 0 & 0 \\
0 & 0 & 0 & 0 & 1 & 0 & 0 & 0 & 0 \\
0 & 0 & 0 & 0 & 0 & 0 & 1 & 0 & 0 \\
\hline
0 & 0 & 1 & 0 & 0 & 0 & 0 & 0 & 0 \\
0 & 0 & 0 & 0 & 0 & 1 & 0 & 0 & 0 \\
0 & 0 & 0 & 0 & 0 & 0 & 0 & 0 & 1 
\end{array} \!\right)\!, 
E_{12}\!=\!\left(\!\begin{array}{ccc|ccc|ccc}
	0 & 0 & 0 & 0 & 0 & 0 & 0 & 0 & 0 \\
	0 & 0 & 0 & 0 & 0 & 0 & 0 & 0 & 0 \\
	0 & 0 & 1 & 0 & -1 & 0 & 1 & 0 & 0 \\
	\hline
	0 & 0 & 0 & 0 & 0 & 0 & 0 & 0 & 0 \\
	0 & 0 & -1 & 0 & 1 & 0 & -1 & 0 & 0 \\
	0 & 0 & 0 & 0 & 0 & 0 & 1 & 0 & 0 \\
	\hline
	0 & 0 & 1 & 0 & -1 & 0 & 1 & 0 & 0 \\
	0 & 0 & 0 & 0 & 0 & 0 & 0 & 0 & 0 \\
	0 & 0 & 0 & 0 & 0 & 0 & 0 & 0 & 0 
\end{array} \!\right)\!,
\end{align}

Choosing accordingly the basis ordering, the two-site IRF density matrix can be simply written as,
\eq
D_2^{\text{IRF}}(\lambda_1,\lambda_2)= \frac{1}{3} D_2^{BT}(\lambda_1,\lambda_2) \oplus  \frac{1}{3} D_2^{BT}(\lambda_1,\lambda_2)\oplus  \frac{1}{3} D_2^{BT}(\lambda_1,\lambda_2).
\label{D2IRF}
\en

On the quantum spin chain side, one has to take the partial trace of the $D_2^{\text{IRF}}(\lambda_1,\lambda_2)$, resulting in the one-site density matrix for the Hamiltonian (\ref{Hirf}),
\eq
{\mathbb D}_1(\lambda_1,\lambda_2)=\left( \begin{array}{ccc}
	\frac{1}{3} & \Omega_0(\lambda_1,\lambda_2) & \Omega_0(\lambda_1,\lambda_2) \\
	\Omega_0(\lambda_1,\lambda_2) & \frac{1}{3} & \Omega_0(\lambda_1,\lambda_2) \\
	\Omega_0(\lambda_1,\lambda_2) & \Omega_0(\lambda_1,\lambda_2) & \frac{1}{3}
\label{D1spin1}
\end{array}\right),
\en
where $\Omega_0(\lambda_1,\lambda_2)=\rho_2(\lambda_1,\lambda_2)-\tfrac{1}{3}\rho_3(\lambda_1,\lambda_2)$.

The only non-trivial one-point correlation function in the ground state of (\ref{Hirf}) is obtained in the homogeneous limit $\lambda_1=\lambda_2 \to 0$, such that,
\eq
\langle S_i^x \rangle_{L} = 2\sqrt{2} ~ \Omega_0(0,0).
\label{X}
\en

Therefore, in order to evaluate the above correlation, one needs first to obtain $\rho_i(0,0)$ \cite{KNS2013}, which can be written as,
\begin{align}
	&\rho_1(0,0) = \tfrac{-1}{90 (3 + 2 \omega(\frac{1}{2}, -\frac{1}{2}))^2}\Big\{-90 
	+ 96 \Big[\omega(-\tfrac{1}{2}, -\tfrac{1}{2}) + \omega(\tfrac{1}{2},\tfrac{1}{2}) - \omega(-\tfrac{1}{2}, \tfrac{1}{2})\Big] - 216 \omega(\tfrac{1}{2}, -\tfrac{1}{2}) \nonumber\\
	&- 128 \omega(-\tfrac{1}{2}, \tfrac{1}{2}) \omega(\tfrac{1}{2}, -\tfrac{1}{2}) - 
	40 \omega(\tfrac{1}{2}, -\tfrac{1}{2})^2 
	+ 192\omega(-\tfrac{1}{2},-\tfrac{1}{2})\omega(\tfrac{1}{2},\tfrac{1}{2}) \nonumber \\
	&+ 48 \Big[\omega^{(1,0)}(-\tfrac{1}{2}, \tfrac{1}{2}) - 
	\omega^{(1,0)}(\tfrac{1}{2}, -\tfrac{1}{2}) + \omega^{(1,0)}(-\tfrac{1}{2}, \tfrac{1}{2}) \omega^{(1,0)}(\tfrac{1}{2}, 
	-\tfrac{1}{2})\Big]  \nonumber\\
	&+ 36 \Big[\omega^{(2,0)}(-\tfrac{1}{2}, \tfrac{1}{2}) + \omega^{(2,0)}(\tfrac{1}{2}, -\tfrac{1}{2})\Big] + 24 \Big[ \omega(\tfrac{1}{2}, -\tfrac{1}{2}) \omega^{(2,0)}(-\tfrac{1}{2}, \tfrac{1}{2}) + \omega(-\tfrac{1}{2}, \tfrac{1}{2}) 
	\omega^{(2,0)}(\tfrac{1}{2}, -\tfrac{1}{2}) \Big] \nonumber \\
	&- 36 \Big[\omega^{(0,2)}(\tfrac{1}{2}, -\tfrac{1}{2}) + 
	2 \omega^{(1,1)}(\tfrac{1}{2}, -\tfrac{1}{2}) + \omega^{(2,0)}(\tfrac{1}{2}, 
	-\tfrac{1}{2})\Big] \nonumber \\
	&- 24 \omega(\tfrac{1}{2}, -\tfrac{1}{2}) \Big[\omega^{(0,2)}(\tfrac{1}{2}, -\tfrac{1}{2}) + 
	2 \omega^{(1,1)}(\tfrac{1}{2}, -\tfrac{1}{2}) + \omega^{(2,0)}(\tfrac{1}{2}, -\tfrac{1}{2})\Big] \Big\}, \label{rho100}
\end{align}
\begin{align}
	&\rho_2(0,0) = \tfrac{-2}{45 (3 + 2 \omega(\tfrac{1}{2}, -\tfrac{1}{2}))^2} \Big\{ 
	96 \Big[\omega(\tfrac{1}{2}, -\tfrac{1}{2}) 
	- \omega(-\tfrac{1}{2}, -\tfrac{1}{2}) 
	- \omega(\tfrac{1}{2}, \tfrac{1}{2}) \Big] \nonumber\\
	&+ 36 \omega(-\tfrac{1}{2}, \tfrac{1}{2}) 
	+ 68 \omega(-\tfrac{1}{2}, \tfrac{1}{2}) \omega(\tfrac{1}{2}, -\tfrac{1}{2}) 
	+ 20 \omega(\tfrac{1}{2}, -\tfrac{1}{2})^2 
	- 112\omega(-\tfrac{1}{2},-\tfrac{1}{2})\omega(\tfrac{1}{2},\tfrac{1}{2}) \nonumber \\ 
	&+ 48 \Big[\omega^{(1,0)}(\tfrac{1}{2}, -\tfrac{1}{2}) 
	- \omega^{(1,0)}(-\tfrac{1}{2}, \tfrac{1}{2}) \Big]
	-28\omega^{(1,0)}(-\tfrac{1}{2},\tfrac{1}{2})\omega^{(1,0)}(\tfrac{1}{2}, -\tfrac{1}{2}) \nonumber\\
	&- 21 \Big[ \omega^{(2,0)}(-\tfrac{1}{2}, \tfrac{1}{2}) 
	- \omega^{(2,0)}(\tfrac{1}{2}, -\tfrac{1}{2}) \Big]
	- 14 \Big[\omega(\tfrac{1}{2}, -\tfrac{1}{2}) \omega^{(2,0)}(-\tfrac{1}{2}, \tfrac{1}{2}) 
	- \omega(-\tfrac{1}{2}, \tfrac{1}{2}) \omega^{(2,0)}(\tfrac{1}{2},-\tfrac{1}{2}) \Big] \nonumber\\
	&+ 21 \Big[\omega^{(0,2)}(\tfrac{1}{2}, -\tfrac{1}{2}) + 2 \omega^{(1,1)}(\tfrac{1}{2}, -\tfrac{1}{2}) + \omega^{(2,0)}(\tfrac{1}{2}, -\tfrac{1}{2})\Big] \nonumber \\
	&+ 14 \omega(\tfrac{1}{2}, -\tfrac{1}{2}) \Big[\omega^{(0,2)}(\tfrac{1}{2}, -\tfrac{1}{2}) + 2 \omega^{(1,1)}(\tfrac{1}{2}, -\tfrac{1}{2}) +\omega^{(2,0)}(\tfrac{1}{2}, -\tfrac{1}{2})\Big] \Big\},
\end{align}
\begin{align}
	&\rho_3(0,0) = \tfrac{-2}{45 (3 + 2 \omega(\tfrac{1}{2}, -\tfrac{1}{2}))^2}\Big\{
	24 \Big[ \omega(-\tfrac{1}{2}, -\tfrac{1}{2}) 
	- \omega(\tfrac{1}{2}, -\tfrac{1}{2}) 
	+ \omega(\tfrac{1}{2}, \tfrac{1}{2}) \Big] \nonumber\\ 
	&+ 36 \omega(-\tfrac{1}{2}, \tfrac{1}{2}) 
	+ 28 \omega(-\tfrac{1}{2}, \tfrac{1}{2}) \omega(\tfrac{1}{2}, -\tfrac{1}{2}) 
	- 20 \omega(\tfrac{1}{2}, -\tfrac{1}{2})^2 
	- 32 \omega(-\tfrac{1}{2}, -\tfrac{1}{2}) \omega(\tfrac{1}{2}, \tfrac{1}{2}) \nonumber\\
	&+ 12 \Big[\omega^{(1,0)}(-\tfrac{1}{2}, \tfrac{1}{2}) 
	- \omega^{(1,0)}(\tfrac{1}{2}, -\tfrac{1}{2}) \Big]
	- 8 \omega^{(1,0)}(-\tfrac{1}{2}, \tfrac{1}{2}) \omega^{(1,0)}(\tfrac{1}{2}, -\tfrac{1}{2}) \nonumber\\
	& - 6 \Big[\omega^{(2,0)}(-\tfrac{1}{2}, \tfrac{1}{2}) 
	- \omega^{(2,0)}(\tfrac{1}{2}, -\tfrac{1}{2}) \Big]
	- 4 \Big[ \omega(\tfrac{1}{2}, -\tfrac{1}{2}) \omega^{(2,0)}(-\tfrac{1}{2}, \tfrac{1}{2}) 
	- \omega(-\tfrac{1}{2}, \tfrac{1}{2}) \omega^{(2,0)}(\tfrac{1}{2}, -\tfrac{1}{2}) \Big] \nonumber\\
	&+ 6 \Big[\omega^{(0,2)}(\tfrac{1}{2}, -\tfrac{1}{2}) 
	+ 2 \omega^{(1,1)}(\tfrac{1}{2}, -\tfrac{1}{2}) + \omega^{(2,0)}(\tfrac{1}{2}, -\tfrac{1}{2})\Big] \nonumber\\
	&+ 4 \omega(\tfrac{1}{2}, -\tfrac{1}{2}) \Big[\omega^{(0,2)}(\tfrac{1}{2}, -\tfrac{1}{2}) + 2 \omega^{(1,1)}(\tfrac{1}{2}, -\tfrac{1}{2}) + \omega^{(2,0)}(\tfrac{1}{2}, -\tfrac{1}{2})\Big] \Big\},\label{rho300}
\end{align}
where $\omega^{(m,n)}(\pm \tfrac{1}{2}, \pm \tfrac{1}{2})=\partial_{\lambda_1}^m \partial_{\lambda_2}^n \omega(\lambda_1,\lambda_2)|_{\lambda_1=\pm \frac{1}{2},\lambda_2=\pm \frac{1}{2}}$. The function $\omega(\lambda_1,\lambda_2)$ can be evaluated in the thermodynamic limit as well at finite system sizes.

\subsubsection{Solution in the thermodynamic limit}

In the thermodynamic limit ($L\to \infty$), the function $\omega(\lambda_1,\lambda_2)$ can be determined by the following three-point equation,
\bear
\frac{ \omega( \lambda_1 -1, \lambda_2)}{(\lambda -1 )^2-1} +  \frac{2 \omega(\lambda_1, \lambda_2)}{\lambda^2-1} + \frac{ \omega( \lambda_1+1, \lambda_2)}{(\lambda +1 )^2-1} = \frac{3}{(\lambda^2-1)(\lambda^2-4)},
\ear
where $\lambda=\lambda_1-\lambda_2$. The solution to this equation can be written as,
\bear
\omega(\lambda,\mu)&=&\frac{1}{2}+\frac{(\lambda-\mu)^2-1}{2} \Big\{\psi^{(0)}(-\frac{1}{2}(\lambda-\mu)) + \psi^{(0)}(\frac{1}{2}(\lambda-\mu))  \\
&-& \psi^{(0)}(\frac{1}{2}-\frac{1}{2}(\lambda-\mu)) -\psi^{(0)}(\frac{1}{2}+\frac{1}{2}(\lambda-\mu))  -\frac{\pi(\lambda-\mu)}{\sin(\pi(\lambda-\mu))} \Big\}. \nonumber
\ear

Therefore, the expansion coefficients of the density matrix $\rho_i(0,0)$ in the thermodynamic limit are given by \cite{KNS2013},
\bear
\rho_1(0,0)= \frac{1}{2}-\frac{2}{45}\pi^2, \quad 
\rho_2(0,0)= -\frac{19}{18} + \frac{14}{135} \pi^2, \quad
\rho_3(0,0)= -\frac{1}{9} + \frac{4}{135} \pi^2, 
\ear
which implies that for $L\to \infty$,
\bear
\langle S_i^x \rangle_{\infty}=2\sqrt2\left(-\frac{55}{27}+\frac{76}{405}\pi^2\right)=-0.2615773595375157..., 
\ear

\subsection{Computation of the three-site density matrix}\label{3siteDM}

The three-site density matrix of the Babujian-Taktahjian model can conveniently be written in terms of  identity, permutation and monoid operators acting on three different sites, such that
as,
\bear
D_3^{BT}(\lambda_1,\lambda_2,\lambda_3)=\sum_{k=1}^{15}\rho_k^{(3)} \check{P}_k,
\ear
where $\check{P}_k$ for $k=1,\dots,15$ can be taken from the following ordered set of linearly independent operators,
\bear
&&\{\check{P}_k\}_{k=1}^{15}=\{ I_{27}, ~P_{12}, ~P_{23}, ~P_{23}P_{12}, ~P_{12}P_{23}, ~P_{13}, ~E_{23}, ~E_{12}, ~E_{13},
\nonumber \\
&& ~E_{12}E_{23}, ~E_{23}E_{12},  ~P_{23} E_{12}, ~P_{12}E_{23}, ~E_{23}P_{12}, ~E_{12}P_{23}  \}, \nonumber
\ear
where $\rho_k^{(3)}=\rho_k^{(3)}(\lambda_1,\lambda_2,\lambda_3)$ are functions that are determined by normalization of the density matrix, by the functional equations and by the density matrix symmetries, which were computed explicitly in the thermodynamic limit \cite{KNS2013}. The final results in the homogeneous limit are given by \cite{KNS2013},
\bear
\rho_1^{(3)}(0,0,0)&=&\tfrac{1879}{54} - \tfrac{13988}{675} \pi^2 + \tfrac{424}{135} \pi^4 - \tfrac{90368}{637875} \pi^6, \nonumber \\ 
\rho_2^{(3)}(0,0,0)&=&\tfrac{1043}{270} - \tfrac{1906}{675} \pi^2 + \tfrac{296}{675} \pi^4 - \tfrac{12416}{637875} \pi^6, \nonumber \\
\rho_4^{(3)}(0,0,0)&=&-\tfrac{251}{18} + \tfrac{1966}{225} \pi^2 - \tfrac{4}{3} \pi^4 + \tfrac{12736}{212625} \pi^6, \nonumber \\
\rho_6^{(3)}(0,0,0)&=&-\tfrac{13021}{270} + \tfrac{788}{27} \pi^2 - \tfrac{2992}{675} \pi^4 + \tfrac{25472}{127575} \pi^6, \nonumber \\
\rho_7^{(3)}(0,0,0)&=&-\tfrac{2917}{135} + \tfrac{8756}{675} \pi^2 - \tfrac{1328}{675} \pi^4 + \tfrac{56576}{637875} \pi^6, \\
\rho_9^{(3)}(0,0,0)&=&-\tfrac{5119}{135} + \tfrac{5128}{225} \pi^2 - \tfrac{2336}{675} \pi^4 + \tfrac{99584}{637875} \pi^6, \nonumber \\
\rho_{10}^{(3)}(0,0,0)&=& \tfrac{371}{15} - \tfrac{2024}{135} \pi^2 + \tfrac{512}{225} \pi^4 - \tfrac{4352}{42525} \pi^6, \nonumber \\
\rho_{12}^{(3)}(0,0,0)&=&\tfrac{343}{45} - \tfrac{3004}{675} \pi^2 + \tfrac{152}{225} \pi^4 - \tfrac{2176}{70875} \pi^6, \nonumber \\
\rho_3^{(3)}(0,0,0)&=&\rho_2^{(3)}(0,0,0), \quad \rho_5^{(3)}(0,0,0)=\rho_4^{(3)}(0,0,0), \nonumber \\
\rho_8^{(3)}(0,0,0)&=&\rho_7^{(3)}(0,0,0), \quad \rho_{11}^{(3)}(0,0,0)=\rho_{10}^{(3)}(0,0,0), \nonumber \\
\rho_{13}^{(3)}(0,0,0)&=&\rho_{14}^{(3)}(0,0,0)=\rho_{15}^{(3)}(0,0,0)=\rho_{12}^{(3)}(0,0,0). \nonumber 
\ear

Again choosing the basis ordering accordingly, the three-site IRF density matrix can be written as,
\eq
D_3^{\text{IRF}}(\lambda_1,\lambda_2,\lambda_3)= \tfrac{1}{3} D_3^{BT}(\lambda_1,\lambda_2,\lambda_3) \oplus  \tfrac{1}{3} D_3^{BT}(\lambda_1,\lambda_2,\lambda_3)\oplus  \tfrac{1}{3} D_3^{BT}(\lambda_1,\lambda_2,\lambda_3).
\label{D3IRF}
\en

On the quantum spin chain side, one has to take the partial trace on  $D_3^{\text{IRF}}(\lambda_1,\lambda_2,\lambda_3)$, which results in the two-site density matrix of the Hamiltonian (\ref{Hirf}),
\eq
{\mathbb D}_2(\lambda_1,\lambda_2,\lambda_3)=\left( \begin{array}{ccc|ccc|ccc}
\tfrac{1}{9} & \Omega_1 & \Omega_1 & \Omega_1 & \Omega_2 & \Omega_3 & \Omega_1 & \Omega_3 & \Omega_2\\ 
\Omega_1 & \tfrac{1}{9} & \Omega_4 & \Omega_5 & \Omega_1 & \Omega_6 & \Omega_6 & \Omega_4 & \Omega_3\\ 
\Omega_1 & \Omega_4 & \tfrac{1}{9} & \Omega_6 & \Omega_3 & \Omega_4 & \Omega_5 & \Omega_6 & \Omega_1\\ 
\hline
\Omega_1 & \Omega_5 & \Omega_6 & \tfrac{1}{9} & \Omega_1 & \Omega_4 & \Omega_4 & \Omega_6 & \Omega_3\\ 
\Omega_2 & \Omega_1 & \Omega_3 & \Omega_1 & \tfrac{1}{9} & \Omega_1 & \Omega_3 & \Omega_1 & \Omega_2\\ 
\Omega_3 & \Omega_6 & \Omega_4 & \Omega_4 & \Omega_1 & \tfrac{1}{9} & \Omega_6 & \Omega_5 & \Omega_1\\ 
\hline
\Omega_1 & \Omega_6 & \Omega_5 & \Omega_4 & \Omega_3 & \Omega_6 & \tfrac{1}{9} & \Omega_4 & \Omega_1\\ 
\Omega_3 & \Omega_4 & \Omega_6 & \Omega_6 & \Omega_1 & \Omega_5 & \Omega_4 & \tfrac{1}{9} & \Omega_1\\ 
\Omega_2 & \Omega_3 & \Omega_1 & \Omega_3 & \Omega_2 & \Omega_1 & \Omega_1 & \Omega_1 & \tfrac{1}{9}
\label{d3}
\end{array}\right),
\en
whose matrix elements in the thermodynamic limit ($L\to \infty$) are given by,
\bear
\Omega_1(0,0,0)&=&-\tfrac{17}{54} + \tfrac{2}{81} \pi^2 =-0.0711209... \nonumber \\ 
\Omega_2(0,0,0)&=&-\tfrac{31231}{810} + \tfrac{47288}{2025} \pi^2 - \tfrac{7192}{2025} \pi^4 + \tfrac{306688}{1913625}\pi^6 =0.038063... \nonumber\\
\Omega_3(0,0,0)&=& -\tfrac{3481}{135} + \tfrac{32036} {2025} \pi^2 - \tfrac{1624}{675} \pi^4 + \tfrac{9856}{91125} \pi^6 =-0.021653... \\
\Omega_4(0,0,0)&=& -\tfrac{7}{18} + \tfrac{2}{45} \pi^2 =0.0497602... \nonumber\\
\Omega_5(0,0,0)&=& 
-\tfrac{401}{135} + \tfrac{3572}{2025} \pi^2 - \tfrac{184}{675}\pi^4 + \tfrac{128}{10125} \pi^6 = 0.0399862...\nonumber\\
\Omega_6(0,0,0)&=& 
-\tfrac{22999}{810} + \tfrac{35272}{2025} \pi^2 - \tfrac{5368}{2025} \pi^4 + \tfrac{228352}{1913625} \pi^6 = 0.0214858... \nonumber
\ear

We use the two-site density matrix (\ref{d3}) to compute some non-trivial correlators in the ground state of the Hamiltonian (\ref{Hirf}) in the thermodynamic limit, which results,
\bear
\langle S_1^x \rangle &=& \sqrt{2}\left(-\tfrac{55}{27} + \tfrac{76}{405}\pi^2\right)=-0.261577..., \nonumber \\   
\langle S_1^x (S_2^z)^2 \rangle &=& \sqrt{2}\left(-\tfrac{38}{27} + \tfrac{56}{405}\pi^2\right)=-0.0604171..., \nonumber \\   
\langle S_1^x S_2^x \rangle &=& -\tfrac{77522}{405} + \tfrac{236336}{2025}\pi^2 - \tfrac{35968}{2025} \pi^4 + \tfrac{
1532416}{1913625} \pi^6=0.155765..., \nonumber \\   
\langle S_1^y S_2^y \rangle &=& \tfrac{30938}{405} - \tfrac{93904}{2025} \pi^2 + \tfrac{14272}{2025} \pi^4 - \tfrac{607744}{1913625} \pi^6=-0.0824327...,  \\   
\langle (S_1^x S_2^x)^2 \rangle &=& -\tfrac{35197}{1620} + \tfrac{1714}{135} \pi^2 - \tfrac{3872}{2025}\pi^4 + \tfrac{
33088}{382725}  \pi^6=0.440748...,, \nonumber \\   
\langle (S_1^y S_2^y)^2 \rangle &=& -\tfrac{30637}{1620} + \tfrac{1006}{81} \pi^2 - \tfrac{3872}{2025} \pi^4 + \tfrac{
33088}{382725}  \pi^6=0.526191... \nonumber 
\ear

\section{Finite system size results}\label{finite}

\subsection{Integral equations}

Similarly to the spin-$1/2$ \cite{TAVARES2023}, the physical properties of the spin-$1$ IRF model and its associated spin chain is obtained from the leading eigenvalue of the vertex model transfer matrix with periodic boundary conditions. Besides, the leading eigenvalues can be determined via non-linear integral equations \cite{SUZUKI99}.

More specifically, the leading eigenvalue $\Lambda_0^{[1]}(\lambda)$ of the row-to-row transfer matrix of the nineteen-vertex (\ref{R19v}) for finite system sizes is given by \cite{KNS2013},
\eq
\ln\left[\frac{\Lambda_0^{[1]}(\im x-\tfrac{1}{2})}{(\im x+\tfrac{1}{2})^L}\right]= L ~ e(x+\tfrac{\im}{2}) + \left( K \ast \ln{B \bar{B}}\right)(x), \label{NLIEeigenvalue}
\en
where 
\eq
e(x)=\log\left[\frac{ \Gamma\left(\frac{3}{2}- \frac{\im x}{2}\right)\Gamma\left(\frac{1}{2}+ \frac{\im x}{2}\right)}{ \Gamma\left(\frac{3}{2}+ \frac{\im x}{2}\right)\Gamma\left(\frac{1}{2}- \frac{\im x}{2}\right)}\right],
\en
and
\eq
K(x)=\frac{\pi}{\cosh{ \pi x}}.
\en
The symbol $\ast$ denotes convolution $(f*g)(x)=\frac{1}{2 \pi}\int_{-\infty}^{\infty} f(x-y)g(y)dy$.

The auxiliary functions $y(x)$, $b(x)$ and $\bar{b}(x)$ and its simply related functions $Y(x)=y(x)+1$, $B(x)=b(x)+1$ and $\bar{B}(x)=\bar{b}(x)+1$ are solutions of the following set
of non-linear integral equations \cite{KNS2013},
\begin{subequations}
	\begin{align}
		\ln{y(x)}&=  -\im \frac{\pi \delta}{\cosh(\pi(x-\im\mu))}  +  \left( \!K\!\ast\! \ln{B }\right)\!(x) + \left(\! K\!\ast\! \ln{\bar{B}}\right)\!(x), \label{NLIE1}\\
		\ln{b(x)}&=  L \ln(\tanh(\tfrac{\pi x}{2}))  +  \left( \!K\!\ast\! \ln{Y}\right)\!(x) +  \left( \!F\!\ast\! \ln{B}\right)\!(x) -
		\left(\! F\!\ast\! \ln{\bar{B}}\right)\!(x+\im), \label{NLIE2}\\
		\ln{\bar{b}(x)}&= L \ln(\tanh(\tfrac{\pi x}{2})) +  \left( \!K\!\ast\! \ln{Y}\right)\!(x)  - \left(\! F \!\ast\! \ln{B}\right)\!(x-\im) + \left(\! F\!\ast\! \ln{\bar{B}}\right)\!(x), \label{NLIE3}
	\end{align}
\end{subequations}
where $\delta$ is a parameter for the construction of two-site correlators and the remaining Kernel function is given by
\eq
F(x)=\int_{-\infty}^{\infty}\frac{e^{-|k|/2+\im k x}}{2 \cosh(\tfrac{k}{2})} dk.
\en

In addition, the two-site correlation function $\omega(\lambda,\mu)$ is given by \cite{KNS2013},
\eq
\omega(\lambda,\mu)=\frac{1}{2}+\frac{(\lambda-\mu)^2 -1}{\im}\Psi(\lambda,\mu).
\label{omegapsi}
\en
This $\Psi(\lambda,\mu)$ function is ingeniously defined from slightly modified transfer matrix \cite{KNS2013} with spin-$1/2$ in the auxiliary space and the insertion of two additional spin-$1/2$ along the quantum direction carrying a $\delta$ parameter (see more details below), such that
\bear
\Psi(\lambda,\mu) &=& \partial_{\delta}  \log\Big[ \tfrac{\Lambda_0^{[\frac{1}{2}]}(\lambda;\mu,\delta)}{(1+\lambda-\mu)(1-(\lambda-\mu-\delta))}\Big] \Big\vert_{\delta =0}.
\label{PsiDelta}
\ear

By the use of fusion relations between the eigenvalues, the above function can be written as \cite{KNS2013},
\bear
\Psi(\lambda-\tfrac{1}{2},\mu-\tfrac{1}{2})\!&=&\! 2 F(\im\lambda -\im\mu)+\! \int_{-\infty}^{\infty}\!\!\frac{1}{\cosh(\pi(\im\lambda-\!x))} \!\left[\! \frac{g_{y}(x;\mu)}{1+y^{-1}(x)}\!\right]\!dx,
\label{sing}
\ear
where the additional auxiliary functions $g_{y}(x;\mu)$, $g_{b}(x;\mu)$ and $g_{\bar{b}}(x;\mu)$ are solution of the following set of linear integral equations \cite{KNS2013},
\begin{subequations}
	\begin{align}
		g_{y}(x;\mu)&=   -\im\frac{\pi}{\cosh(\pi(x-\im\mu))}  +  \left( \!K\!\ast\! \frac{g_{b}(x;\mu)}{1+b^{-1}} \right)\!(x) + 
		\left(\! K\!\ast\! \frac{g_{\bar{b}}(x;\mu)}{1+\bar{b}^{-1}}\right)\!(x), \label{NLIEG1}\\
		g_{b}(x;\mu)&= \left( \!K\!\ast\! \frac{g_{y}(x;\mu)}{1+y^{-1}} \right)\!(x) + \left( \!F\!\ast\! \frac{g_{b}(x;\mu)}{1+b^{-1}} \right)\!(x) - \left(\! F\!\ast\!\frac{g_{\bar{b}}(x;\mu)}{1+\bar{b}^{-1}}\right)\!(x+\im), \label{NLIEG2}\\
		g_{\bar{b}}(x;\mu)&=   \left( \!K\!\ast\! \frac{g_{y}(x;\mu)}{1+y^{-1}} \right)\!(x) - \left( \!F\!\ast\! \frac{g_{b}(x;\mu)}{1+b^{-1}} \right)\!(x-\im) + \left(\! F\!\ast\!\frac{g_{\bar{b}}(x;\mu)}{1+\bar{b}^{-1}}\right)\!(x).
		\label{NLIEG3}
	\end{align}
\end{subequations}

By solving the above integral equations, one can determine the physical properties of the model as well as the two-site $\omega$-function evaluated at the fusion points $\pm \frac{1}{2}$, such that $\omega(\pm \frac{1}{2},\pm \frac{1}{2})$. Nevertheless, in order to take the homogeneous limit of the $\rho_i(0,0)$ (\ref{rho100}-\ref{rho300}), we also need to compute derivatives of the $\omega$-function evaluated at the fusion points. However, this requires a singular limit, which implies in the existence of a singularity inside the convolution in (\ref{sing}), this makes the evaluation of the $\omega$-derivatives very intricate. In order to circumvent this issue, we take a different route and evaluate the derivatives $\omega^{(m,n)}(\pm \tfrac{1}{2}, \pm \tfrac{1}{2})$ directly from the definition (\ref{PsiDelta}) via the solution of the Bethe ansatz equation. This procedure will be described in what follows.

\subsection{Bethe ansatz solution}

In order to make clear the subtleties in the evaluation of the derivatives $\omega^{(m,n)}(\pm \tfrac{1}{2}, \pm \tfrac{1}{2})$ at the special fusion points, we have to give more details about the context where the function $\omega(\lambda,\mu)$ is defined. This special function is related to the expected value of the permutation operator, with support on a local space formed by two spin-$1/2$. In the vertex language, it has been shown \cite{KNS2013} that this expectation can be obtained from the transfer matrix,
\eq
T(\lambda;\mu,\delta) =\tr_A\left[ R_{A L+2}^{[\frac{1}{2},\frac{1}{2}]} (\lambda-\mu) {R_{L+1 A}^{[\frac{1}{2} \frac{1}{2}]}}^{t_1} (\mu-\lambda+\delta)  R_{A L}^{[\frac{1}{2},1]}(\lambda-\tfrac{1}{2}) \ldots R_{A 1}^{[\frac{1}{2},1]} (\lambda-\tfrac{1}{2})\right], \label{COSTURA}
\en
through the logarithmic derivative of its largest eigenvalue $\Lambda_0^{[\frac{1}{2}]}(\lambda; \mu,\delta)$ as defined in Eq.(\ref{PsiDelta}).
Here the $R$-operators are the fundamental and a spin-$1$ solution of the Yang-Baxter equation as indicated by the super-indices, according to the conventions in \cite{RIBEIRO2008}.
For simplicity, we use $\lambda = -\im x$ and $\mu = - \im y$. In terms of the algebraic Bethe ansatz, the dominant eigenvector is found after the action of $L+1$ creation operators, usually denoted as $\tilde{B}(x_j)$, over the reference state $\vert 1\ldots 1 \rangle \otimes \vert \downarrow \uparrow\rangle$. 
Generally, the eigenvector  is obtained by the action of $n$ such creation operators, resulting in the expression for the eigenvalue
\eq
\Lambda^{[\frac{1}{2}]}(x; y,\delta) = \lambda_1(x) +\lambda_2(x), \label{EIG}
\en 
where
\eq
\lambda_1(x)=\Phi(x+\im) \phi(x+\im) \phi(x-\delta)\frac{Q(x-\im)}{Q(x)},
\en
\eq
\lambda_2(x)=\Phi(x-\im) \phi(x) \phi(x-\im -\delta)\frac{Q(x+\im)}{Q(x)},
\en
and
\eq
Q(x) = \prod_{j=1}^{n} x-x_j, \qquad \Phi(x) = x^L, \qquad \phi(x) = x-y.
\en
The Bethe roots $x_j$ must satisfy the Bethe equations in the form,
\eq
\lim_{x \to x_k} - \frac{\lambda_1(x)}{\lambda_2(x)}=1, \qquad k=1,\ldots, n. \label{BAE}
\en
The special texture of the transfer-matrix (\ref{COSTURA}) implies the existence of factorized eigenvectors in the limit $\delta \to 0$. This is so because of the unitarity condition satisfied by the fundamental $R$-matrix, see Figure \ref{UNIT}.

\begin{figure}[bht]
	\begin{tikzpicture}[line width=0.5pt]
		\begin{scope}
			\draw (-0.5,0)--(5.5,0);
			\draw (0,-1.4)--(0,0.9);
			\draw (1,-1.4)--(1,0.9);
			\foreach \x in {2,3,4,5}
			\draw[thick,double] (\x,-1.4)--(\x,1.4);
			\draw (0,0.9) arc[start angle=180,end angle=0,radius=0.5];
			\draw (0.18,0) arc[start angle=0,end angle=90,radius=0.18];
			\draw (1.18,0) arc[start angle=0,end angle=-90,radius=0.18];
			\foreach \x in {2,3,4,5}
			\draw (\x+0.18,0) arc[start angle=0,end angle=90,radius=0.18];
		\end{scope}
		\node at (7,0) {$=$};
		\begin{scope}[xshift=9cm]
			\draw (-0.5,0)--(5.5,0);
			\draw (0,-0.9)--(0,-1.4);
			\draw (1,-0.9)--(1,-1.4);
			\draw (1,-0.9) arc[start angle=0,end angle=180,radius=0.5];
			\foreach \x in {2,3,4,5}
			\draw[thick,double] (\x,-1.4)--(\x,1.4);
			\foreach \x in {2,3,4,5}
			\draw (\x+0.18,0) arc[start angle=0,end angle=90,radius=0.18];
		\end{scope}
	\end{tikzpicture}
	\caption{Factorization of the eigenvector at $\delta =0$. We depict single lines for the spin-$1/2$ objects and double lines for spin-$1$.}
	\label{UNIT}
\end{figure}

At this point, a contraction between the two last columns (from right to left), i.e. a vector of type $\vert \Psi \rangle \otimes \left( \vert \uparrow \uparrow \rangle+\vert \downarrow \downarrow \rangle\right)$, leads to the following equation
\eq
T(x;y,0)  \vert \Psi \rangle \otimes \left( \vert \uparrow \uparrow \rangle+\vert \downarrow \downarrow \rangle\right) = (1+{(x-y)}^2) (t(x)   \vert \Psi \rangle)  \otimes \left( \vert \uparrow \uparrow \rangle+\vert \downarrow \downarrow \rangle\right), \label{fact}
\en    
where 
\eq
t(x) =\tr_A  R_{A L}^{[\frac{1}{2},1]} (-\im x-1/2) \ldots R_{A 1}^{[\frac{1}{2},1]} (-\im x-1/2), \label{HOMO}
\en
acts solely in the homogeneous spin-1 Hilbert space. Such a transfer matrix belongs to the fusion hierarchy of the nineteen-vertex model, i.e.
\eq
t(x-\im/2) t(x+\im/2) = T(x) + \Phi(x+3 \im/2) \Phi(x-3 \im/2) \mbox{Id},
\en  
with $T(x)$ the vertex counterpart of $T_{IRF}(x)$.  As the  map (\ref{MAP}) indicates, we can extract spectral properties of $T_{IRF}(x)$ from $T(x)$. For the dominant eigenvalue, we stick to the twist sector of $\phi =0$ and the eigenvalue expression for $t(x)$ is given as
\eq
\Lambda^{[\frac{1}{2}]}(x) = \lambda_1^{(0)}(x) + \lambda_2^{(0)}(x),
\en
where
\eq
\lambda_1^{(0)}(x)=\Phi(x+\im)\frac{Q(x-\im)}{Q(x)},
\en
\eq
\lambda_2^{(0)}(x)=\Phi(x-\im) \frac{Q(x+\im)}{Q(x)},
\en
with Bethe equations
\eq
\lim_{x \to x_k} - \frac{\lambda_1^{(0)}(x)}{\lambda_2^{(0)}(x)}=1, \qquad k=1,\ldots, n.
\en
To understand how the eigenvectors in (\ref{fact}) possibly appear from an $n$-particle state created over the vacuum $ \vert 1\ldots 1 \rangle \otimes \vert \downarrow \uparrow\rangle$, we perform a two-side model decomposition, namely we write operators of the Yang-Baxter algebra as
\eq
{\cal T}  = 
\left(\begin{array}{c|c}
	A_{L+2} A_{L+1}+B_{L+2} B_{L+1} & A_{L+2} C_{L+1}+B_{L+2} D_{L+1} \\
	\hline
	C_{L+2} A_{L+1}+D_{L+2} B_{L+1} & C_{L+2} C_{L+1}+D_{L+2} D_{L+1}
\end{array}\right)
\cdot
\left(\begin{array}{c|c}
	A & B \\
	\hline
	C & D
\end{array}\right),
\en
where the operators without indices carry parameter $x$ and act on the spin-1 Hilbert space, those acting locally at spin-$1/2$ sites $L+1$ and $L+2$ carry parameters $y-x$ and $x-y$, respectively. Therefore, for the selected reference state the creation operator is
\eq
\tilde{B}=(A_{L+2} A_{L+1}+B_{L+2} B_{L+1}) B+ (A_{L+2} C_{L+1}+B_{L+2} D_{L+1}) D.
\en
The first term on the right hand side flips a spin of the bulk spin-$1$ space, while it is stable on the local spin-$1/2$ factors. On the contrary, the second term is diagonal on the bulk while it generates a linear combination of polarized spins on the spin-$1/2$ factor. A single application of the creation operator on a generic state of type $\vert \Psi_0 \rangle \otimes \vert \downarrow \uparrow\rangle$ leads to,
\begin{multline}
	\tilde{B}(x) \vert \Psi_0 \rangle \otimes \vert \downarrow \uparrow\rangle = 
	(1-\im (x-y)) (-\im (y-x)) (B\vert \Psi_0 \rangle) \otimes \vert \downarrow \uparrow\rangle+\\
	(1-\im (x-y))(D \vert \Psi_0 \rangle) \otimes \vert \uparrow \uparrow\rangle+
	(1+\im (x-y)) (D \vert \Psi_0 \rangle) \otimes \vert \downarrow \downarrow\rangle, \label{p0}
\end{multline}
where a non-trivial superposition among polarized and non-polarized spins on factor $1/2$ appears. As for states of the form 
$\vert \Psi_0 \rangle \otimes \vert \uparrow \uparrow\rangle$ and $\vert \Psi_0 \rangle \otimes \vert \downarrow \downarrow\rangle$, we find
\begin{multline}
	\tilde{B}(x) \vert \Psi_0 \rangle \otimes \vert \uparrow \uparrow\rangle = 
	(1-\im (x-y)) (1-\im (y-x)) (B\vert \Psi_0 \rangle) \otimes \vert \uparrow \uparrow\rangle+\\
	1 (B \vert \Psi_0 \rangle) \otimes \vert \downarrow \downarrow\rangle+
	(\im (x-y)) (D \vert \Psi_0 \rangle) \otimes \vert \uparrow \downarrow\rangle, \label{p1}
\end{multline}
\begin{multline}
	\tilde{B}(x) \vert \Psi_0 \rangle \otimes \vert \downarrow \downarrow\rangle = 
	(-\im (x-y)) (-\im (y-x)) (B\vert \Psi_0 \rangle) \otimes \vert \downarrow \downarrow\rangle+\\
	(-\im (x-y)) (D \vert \Psi_0 \rangle) \otimes \vert \uparrow \downarrow\rangle. \label{p2}
\end{multline}
The unitarity simplification is manifested when we add (\ref{p1}) to (\ref{p2}), as expected. There, the polarized states have the same amplitude, whereas the non-polarized part is exactly canceled. In other words, the creation operator is stable over the spin-$1/2$ factor of the  state $ \vert \Psi_0 \rangle \otimes \left(\vert \downarrow \downarrow\rangle+\vert \uparrow \uparrow\rangle\right)$, while it creates only on the bulk part of spin-$1$. Having started with a reference state as in (\ref{p0}), to arrive at states of the form $ \vert \Psi \rangle \otimes \left(\vert \downarrow \downarrow\rangle+\vert \uparrow \uparrow\rangle\right)$ requires the annihilation of the first term on the right hand side of (\ref{p0}), while providing equal amplitudes for the remaining terms. This is achieved by setting a single Bethe root to $x_j=y$. Once this condition is met, we can further create an $n$-particle state with generic $x_j$'s,  for which (\ref{p1}-\ref{p2}) guarantee they are of the required form. 

Therefore, at $\delta =0$, we found a relation between the $n+1$-particle state  Bethe ansatz expression for the eigenvalues of $T(x;y,\delta)$ and the $n$-particle state Bethe ansatz expression for the eigenvalues of $t(x)$. As $\delta \to 0 $, a quarter of the states have a Bethe root, say $x_{n+1}$, flowing to $y$, and we have
\eq
\Lambda^{[\frac{1}{2}]}(x;y,0) = \phi(x+\im) \phi(x-\im) \Lambda^{[\frac{1}{2}]}(x).
\en

In terms of analytical properties, the Bethe root at $y$ has a different status compared to the others. It is well known that the Bethe equations can be viewed as conditions that transforms poles in the Bethe expression of the eigenvalues (due to $Q(x)$) into removable singularities.  Cancellation occurs thanks to the sum $\lambda_1(x) +\lambda_2(x)$. For the root at $y$, the singularity is removed, instead, individually, for each $\lambda_{1,2}(x)$, by the cancellation of the pole in $Q(x)$ with a zero in the vacuum expectation, $\phi(x)$. For an effective numerical solution of the Bethe equations this poses severe difficulties. At $\delta=0$, one either finds sub-leading eigenvalues (much smaller than the one we are interested in), or Bethe roots flowing to the infinity, which is equivalent to shifting away from the sector where the dominant eigenvalue is found. We circumvent this numerical problem by the relations found in the previous paragraphs and explicitly using $x_{L+1}=y$. For the largest eigenvalue, the remaining Bethe roots form two-string pattern with imaginary parts close to $\pm \frac{1}{2}$.

However, for our purposes, not only a tractable Bethe expression for the eigenvalue itself is required. We must further obtain the derivative (\ref{PsiDelta}), i.e. we set out to compute
\eq
\partial_{\delta}  \log \Lambda_0^{[\frac{1}{2}]}(x;y,\delta) \Big\vert_{\delta =0} = \frac{1}{\Lambda^{[\frac{1}{2}]}(x;y,0)}\left( \lambda_1(x) \partial_{\delta}  \log \lambda_1(x) \Big\vert_{\delta =0}   +\lambda_2(x)\partial_{\delta}  \log \lambda_2(x) \Big\vert_{\delta =0} \right), \label{EIGDER}
\en
\begin{align}
	\partial_{\delta} \log \lambda_1\Big \vert_{\delta=0} &= -\frac{1}{x-y} +\sum_{j=1}^{L+1}\left[\frac{1}{x-x_j}-\frac{1}{x-x_j-\im}\right]  X_j, 
	\\
	\partial_{\delta} \log \lambda_2\Big \vert_{\delta=0} &= -\frac{1}{x-y-\im} +\sum_{j=1}^{L+1}\left[\frac{1}{x-x_j}-\frac{1}{x-x_j+\im}\right]  X_j,
	\label{EIGDER2}
\end{align}
where we have defined $X_j = \partial_{\delta} x_j \Big \vert_{\delta=0}$. The derivatives of the Bethe roots can be obtained by taking the derivative of the Bethe equations. In other words, once a solution to the original Bethe equations is found, one can define a linear set of equations for the $X_j$'s:
\eq
\left[\tfrac{2 \im L}{1+x_k^2}+\tfrac{2 \im}{1+{(x_k-y)}^2}+\sum_{j \neq k}\tfrac{-2 \im}{1+{(x_j-x_k)}^2}\right] X_k \\+\sum_{j \neq k} \tfrac{2 \im X_j}{1+ {(x_k-x_j)}^2}= \tfrac{1}{x_k-y-\im}-\tfrac{1}{x_k-y}, \label{BETHEDERIVATIVE}
\en
where the index $k$ runs from $1$ to $L$. One can see immediately that the $L+1$-th equation has to be treated differently. In fact, had we started with generic $\delta$ we would have found, 
\begin{multline}
	\Bigg[\tfrac{2 \im L}{1+x_{L+1}^2}+\frac{1}{x_{L+1}-y-\delta-\im}-\tfrac{1}{x_{L+1}-y+\im}+ \tfrac{1}{x_{L+1}-y}-\tfrac{1}{x_{L+1}-y-\delta} +\\ \sum_{j \neq L+1}\tfrac{-2 \im}{1+{(x_j-x_{L+1})}^2}\Bigg] X_{L+1}+\sum_{j \neq L+1} \tfrac{2 \im X_j}{1+ {(x_{L+1}-x_j)}^2} = \tfrac{1}{x_{L+1}-y-\im} -\tfrac{1}{x_{L+1}-y-\delta}, 
\end{multline} 
where the determination of the value $X_{L+1}$ is conditioned to the evaluation of how fast it converges to its limiting value. The singular terms in $y$ can be grouped such that,
\begin{multline}
	\Bigg[\tfrac{2 \im L}{1+x_{L+1}^2}+\tfrac{1}{x_{L+1}-y-\delta-\im}-\tfrac{1}{x_{L+1}-y+\im} + \sum_{j \neq L+1}\tfrac{-2 \im}{1+{(x_j-x_{L+1})}^2}\Bigg] X_{L+1} \\+ \tfrac{1}{\mathfrak{z}(\mathfrak{z}-1)}\tfrac{\mathfrak{z}-X_{L+1}}{\delta} +\sum_{j \neq L+1} \tfrac{2 \im X_j}{1+ {(x_{L+1}-x_j)}^2} = \tfrac{1}{x_{L+1}-y-\im}, 
\end{multline} 
where  $\mathfrak{z} = \frac{x_{L+1}-y}{\delta}$. 

Finiteness of such a term requires that $x_{L+1}  = y + X_{L+1} \delta + \alpha \delta^2 + o(\delta^2) $. Therefore, for generic $\delta$ we have a linear system of $L+1$ equations for the $L+1$ unknowns $X_j$'s, while for the limit $\delta \to 0$, an equation is missing, since this extra equation requires higher order derivatives of the Bethe root $x_{L+1}$. To overcome this difficulty, we have to obtain $X_{L+1}$ independently.  Only then (\ref{BETHEDERIVATIVE}) becomes efficient in determining the other unknowns.  

Now comes the main step in our argumentation. Thanks to integrability, we know that the eigenvalues (\ref{EIG}-\ref{BAE}) are polynomials in the spectral parameter $x$. This is because all matrix elements of the transfer matrix are polynomials and the eigenvectors do not depend on 
$x$. 
Ultimately, they should be  polynomials of $x$ whose coefficients are $y$ and $\delta$ dependent. As a polynomial, there should be no poles in $x$, as stated before. However, the matrix elements of the transfer matrix are also polynomials in variables $y$ and $\delta$. As such, the secular equation determining all of its eigenvalues have coefficients that are polynomials in these variables. It is straightforward to see that all the eigenvalues (solutions to the secular equation) must be continuous functions of $y$ and $\delta$. Furthermore, a pole in  $\partial_{\delta} \log \Lambda (x;y, \delta) {\Big \vert}_{\delta =0} $ could only  exist in the case where the eigenvalue is degenerate. For the largest eigenvalue, this is in contradiction with the first order expansion in $\delta$ for the eigenvalue, see (\ref{EIGDER}-\ref{EIGDER2}), where at $\delta \to 0$ one finds a divergence at $x=y$. The contradiction is resolved if we set $X_{L+1}$ to kill such a pole,
\eq
X_{L+1} = \frac{\lambda_1^{(0)}(y)}{\lambda_1^{(0)}(y)+\lambda_2^{(0)}(y)}.
\en
After some simplifications, we may write the derived set of equations in the form
\eq
\left[ \tfrac{2 \im L}{1+x^2_k}+ \sum_{j\neq k}\tfrac{-2\im}{1+(x_j-x_k)^2} \right] X_k+\sum_{j=1}^L \tfrac{2 \im X_j}{{(x_k-x_j)}^2+1}=\tfrac{1}{x_k-y-\im}-\tfrac{1}{x_k-y}-\tfrac{2 \im X_{L+1}}{1+{(x_k-y)}^2}, \quad k=1,\ldots, L
\label{BAderivatives}
\en
where it is manifested that no divergence exist. Moreover, we may write an expression for the derivative $\partial_{\delta} \log \Lambda^{[\frac{1}{2}]}(x;y, \delta) {\Big \vert}_{\delta =0} $ in terms of:
\begin{align}
	r_1(x) &= (y-x) -\im X_{L+1} +(1+{(x-y)}^2)\sum_{j=1}^L \left[\frac{1}{x-x_j}-\frac{1}{x-x_j-\im}\right] X_j,
	\\
	r_2(x) &= (y-x-\im) +\im X_{L+1} +(1+{(x-y)}^2)\sum_{j=1}^L \left[\frac{1}{x-x_j}-\frac{1}{x-x_j+\im}\right] X_j,
\end{align}
\begin{multline}
	(1+{(x-y)}^2)\partial_{\delta} \log \Lambda^{[\frac{1}{2}]}(x;y, \delta) {\Big \vert}_{\delta =0} = \frac{\lambda_1^{(0)}(x)r_1(x)+\lambda_2^{(0)}(x) r_2(x)}{\lambda_1^{(0)}(x)+\lambda_2^{(0)}(x)}+\\\frac{1}{x-y} \left[\frac{\lambda_2^{(0)}(y) \lambda_1^{(0)}(x)-\lambda_1^{(0)}(y) \lambda_2^{(0)}(x)}{(\lambda_1^{(0)}(x)+\lambda_2^{(0)}(x)) (\lambda_1^{(0)}(y)+\lambda_2^{(0)}(y))}\right].
\end{multline}
This allows obtaining the Bethe roots $x_j$ as well as the $\delta$ derivative of the Bethe roots $X_j$ which are required to evaluate (\ref{omegapsi}) and its higher order derivatives.

\subsection{Numerical computation for finite system sizes}

\begin{table}[b!]
	\begin{center}
		\begin{tabular}{|c|c|c|c|c|c|c|c|}
			\hline
			$L$ & \!\!\! $\omega(\frac{1}{2},\frac{1}{2})$ \!\!\! & \!\!\! $\omega(\frac{-1}{2},\frac{-1}{2})$ \!\!\!& \!\!\!$\omega^{(1,0)}(\frac{-1}{2},\frac{1}{2})$ \!\!\!\!\!& \!\!$\omega^{(1,0)}(\frac{1}{2},\frac{-1}{2})$ \!\!\!\! & \!\!\!$\omega^{(1,1)}(\frac{-1}{2},\frac{1}{2})$\!\!\!\! & \!\!$\omega^{(2,0)}(\frac{-1}{2},\frac{1}{2})$ \!\!\!\!&
			\!\!$\omega^{(2,0)}(\frac{1}{2},\frac{-1}{2})$ \!\!\!\!\\
			\hline
			$4$ & $-0.528108$ & $-0.308870$ & $- \im 0.058870 $ & $\im 0.234261$ & $-0.098630$ & $0.117130$ & $0.080130$  \\
			\hline
			$8$ & $-0.435781$ & $-0.345396$ & $-\im 0.095395$ & $\im 0.175738$ & $-0.114641$ & $0.114541$ & $0.114741$ \\
			\hline
			$16$ & $-0.407188$ & $-0.365572$ & $-\im 0.115572$ & $\im 0.154739$ & $-0.118801$ & $0.115456$ & $0.122146$\\
			\hline
			$32$ & $-0.396035$ & $-0.376006$ & $-\im 0.126006$ & $\im 0.145428$ & $-0.120077$ & $0.117229$ & $0.122923$ \\
			\hline
			$64$ & $-0.391054$ & $-0.381222$ & $-\im 0.131222$ & $\im 0.140902$ & $-0.120487$ & $0.118672$ & $0.122302$\\	
			\hline $\infty$ & $-0.386294$ & $-0.386294$ & $-\im 0.136294$ & $\im 0.136294$ & $-0.120680$ & $0.120680$ & $0.120680$ \\
			\hline
		\end{tabular}
	\end{center}
\caption{Computation of the $\omega(\pm \frac{1}{2},\pm \frac{1}{2})$ via integral equations/Bethe ansatz and the computation of $\omega^{(m,n)}(\pm \tfrac{1}{2}, \pm \tfrac{1}{2})$ via Bethe ansatz.}
\label{tab0}
\end{table}
\begin{table}[h!]
	\begin{center}
		\begin{tabular}{|l|c|c|c|c|c|}
			\hline
			Length & $\rho_1(0,0)$ & $\rho_2(0,0)$ & $\rho_3(0,0)$ & $\Omega_0(0,0)$ & $\langle S_i^x \rangle$  \\
			\hline
			$L=2$ & $0.0000000$ & \hspace{0.3cm}$0.0000000$ & $0.3333333$ & $-0.1111111$ & $-0.3142697$\\
			\hline
			$L=4$ & $0.0513218$ & $-0.0281275$ & $0.2074955$ & $-0.0972927$ & $-0.2751854$  \\
			\hline
			$L=6$ & $0.0572868$ & $-0.0306823$ & $0.1921550$ & $-0.0947341$ & $-0.2679481$ \\
			\hline
			$L=8$ & $0.0591520$ & $-0.0313737$ & $0.1872509$ & $-0.0937908$ & $-0.2652804$ \\
			\hline
			$L=16$ & $0.0608281$ & $-0.0319091$ & $0.1827582$ & $-0.0928284$ & $-0.2625585$\\
			\hline
			$L=32$ & $0.0612231$ & $-0.0320124$ & $0.1816766$ & $-0.0925713$ & $-0.2618312$ \\
			\hline
			$L=64$ & $0.0613193$ & $-0.0320345$ & $0.1814099$ & $-0.0925045$ & $-0.2616421$\\	
			\hline $L\rightarrow\infty$ & $0.0613509$ & $-0.0320410$ & $0.1813216$ & $-0.0924816$ & $-0.2615774$ \\
			\hline
		\end{tabular}
	\end{center}
	\caption{Comparison of numerical results for the density matrix element via exact diagonalization for $L=2,4,6,8$ and via Bethe ansatz solution for $L=4,8,16,32,64$  with the analytical result in the thermodynamic limit.}
	\label{tab1}
\end{table}

The $\omega(\pm \frac{1}{2},\pm \frac{1}{2})$ and its derivatives $\omega^{(m,n)}(\pm \tfrac{1}{2}, \pm \tfrac{1}{2})$ were computed for various system sizes up to $64$ and their values are displayed in Table \ref{tab0}. There are a couple of special cases, such as  $\omega(\frac{1}{2},-\frac{1}{2})=\omega(-\frac{1}{2},\frac{1}{2})=-\tfrac{1}{2}$ and $\omega^{(0,2)}(-\frac{1}{2},\frac{1}{2})=\omega^{(2,0)}(\frac{1}{2},-\frac{1}{2})$.

Having evaluated $\omega$ and its derivatives at the fusion points, we can calculate the homogeneous limit of the $\rho_i(0,0)$ from (\ref{rho100}-\ref{rho300}). This allows us to determine the density matrix (\ref{D2XXX}) and consequently the IRF density matrices (\ref{D2IRF},\ref{D1spin1}), as well as the non-trivial one-point correlator (\ref{X}) for finite system sizes. 

The analytical results in the thermodynamic limit are in very good agreement with results from exact diagonalization for small lattices sizes, up to $8$, as well as with the Bethe ansatz results up to $64$ sites, as shown in Table \ref{tab1}.

\newpage
Again the analytical result for all three-site reduced density matrix elements and some non-trivial correlation functions in the thermodynamic limit are in good agreement with the results obtained in the exact diagonalization, as show in Table \ref{tab2}-\ref{tab3}.

\begin{table}[h!]
	\begin{center}
		\begin{tabular}{|l|r|r|r|r|r|r|}
			\hline
			Length & $\Omega_1(0,0,0)$ & $\Omega_2(0,0,0)$& $\Omega_3(0,0,0)$& $\Omega_4(0,0,0)$& $\Omega_5(0,0,0)$& $\Omega_6(0,0,0)$ \\
			\hline
			$L=2$ & $-0.1111111$ & $0.1111111$ & $-0.1111111$& $0.1111111$& $
			0.1111111$& $0.1111111$ \\
			\hline
			$L=4$ & $-0.0785410$ & $0.0487691$ & $-0.0314164$& $0.0597893$& $0.0501681$& $0.0300174$ \\
			\hline
			$L=6$ & $-0.0742792$ & $0.0423769$ & $-0.0253127$& $0.0538243$& $0.0441147$& $0.0246212$ \\
			\hline
			$L=8$ & $-0.0728749$ & $0.0404029$ & $-0.0235743$& $0.0519591$& $0.0422409$& $0.0231212$ \\
			\hline
			$L\rightarrow\infty$ & $-0.0711209$ & $0.0380638$ & $-0.0216530$ & $0.0497602$ & $0.0399862$ & $0.0214858$    \\
			\hline
		\end{tabular}
	\end{center}
	\caption{Comparison of numerical results for the density matrix elements from exact diagonalization for $L=2,4,6,8$ sites with the analytical result in the thermodynamic limit.}
	\label{tab2}
\end{table}

\begin{table}[h!]
	\begin{center}
		\begin{tabular}{|l|c|c|c|c|c|c|}
			\hline
			Length & $\langle S_1^x \rangle$   & $\langle S_1^x (S_2^z)^2 \rangle$ & $\langle S_1^x S_2^x \rangle$ & $\langle S_1^y S_2^y \rangle$ & $\langle (S_1^x S_2^x)^2 \rangle$ & $\langle (S_1^y S_2^y)^2 \rangle$    \\
			\hline
			$L=2$ & $-0.3142696$ &  $-0.0000000$ & $0.4444444$ & $-0.4444444$  & $0.5555556$ &  $0.5555556$ \\
			\hline
			$L=4$ & $-0.2751853$ &  $-0.0530378$ & $0.1950762$ & $-0.1200694$  & $0.4564096$ &  $0.5314164$ \\
			\hline
			$L=6$ & $-0.2679483$ &  $-0.0578551$ & $0.1716003$ & $-0.0963923$  & $0.4467805$ &  $0.5286000$ \\
			\hline
			$L=8$ & $-0.2652804$ &  $-0.0591589$ & $0.1643812$ & $-0.0897149$  & $0.4439346$ &  $0.5275980$ \\
			\hline
			$L\rightarrow\infty$ & $-0.2615773$ & $-0.0604171$ & $0.1557654$ & $-0.0824327$  & $0.4407481$ & $0.5261908$  \\
			\hline
		\end{tabular}
	\end{center}
	\caption{Comparison of numerical results from exact diagonalization for $L=2, 4, 6, 8$ sites with the analytical result in the thermodynamic limit for correlations.}
	\label{tab3}
\end{table}

\section{Conclusion}\label{CONCLUSION}

We proposed an IRF version of the Zamolodchikov-Fateev model. This produces a new face model made of three copies of this nineteen-vertex model. The spin chain associated to this face model 
can be written as three coupled Zamolodchikov-Fateev spin-$1$ chains with different boundary twists.

In the isotropic limit ($\gamma\to0$), the Zamolodchikov-Fateev spin chain reduces to the Babujian-Taktahjian spin-$1$ chain. In this special case, we managed to compute short-distance correlations of the IRF version of the isotropic nineteen-vertex model and its associated isotropic spin chain. We computed the two and three-site density matrix in the thermodynamic limit. We also computed the density matrix and non-trivial  correlators for finite system sizes up to $64$ sites. Besides, it is worth to note that, in contrast to the spin-$1/2$ version  \cite{TAVARES2024} where the correlation function of the six-vertex IRF spin chain coincides with certain Heisenberg spin chain correlations \cite{TAVARES2024}, here there is no simple relation among the correlators of the two spin-$1$ chains.  

It is worth noting that the Zamolodchikov-Fateev vertex model was generalized by means of fusion \cite{RESHE,BTsvelik} and the universal $R$-matrix \cite{FADDEEV,Bytsko} techniques, which allowed the explicit construction of the spin-$s$ XXZ chain \cite{Bytsko}. Besides that, there also exists an elliptic $21$-vertex model $R$-matrix associated to a spin-$1$ XYZ chain \cite{ZF1980}. However, this $R$-matrix in the current representation produces an IRF spin-$1$ chain with discrete $\mathbb{Z}_2$ symmetry, while we expect that the generalization of (\ref{Hirf2}) should have discrete $\mathbb{Z}_3$ symmetry.

We expect to obtain new spin chains associated to the face version of the spin-$s$ chains\cite{ZF1980,BTsvelik,Bytsko} and high-rank models\cite{RK2019,RIBEIRO2020}  along the same lines as in the cases discussed here. Finally, we also expect that the understanding of the subtleties in the evaluation of the derivatives of the $\omega$ function at the fusion points in the Bethe ansatz level can help in setting up more tractable version of the integral formulation approach.

\section*{Acknowledgments}

GAPR and HBA thank the S\~ao Paulo Research Foundation (FAPESP) for financial support through the grants 2023/03947-0 and 2025/20082-9.

\end{document}